\documentclass[a4paper,fleqn,usenatbib]{mnras}

\usepackage[T1]{fontenc}
\usepackage{ae,aecompl}
\usepackage{footmisc}

\usepackage{graphicx}	
\usepackage{pdflscape}
\usepackage{flushend}

\title[MAXI J1535$-$571 - {\it AstroSat} view]{{\it AstroSat} view of MAXI
J1535$-$571: broadband spectro-temporal features}

\author[Sreehari et al.]{Sreehari H.$^{1,5}$\thanks{E-mail: hjsreehari@gmail.com},
Ravishankar B. T.$^{1,2}$,
Nirmal Iyer$^{3}$,
V. K. Agrawal$^{1}$,
Tilak B. Katoch$^{4}$,
\newauthor Samir Mandal$^{2}$,
Anuj Nandi$^{1}$ \\
\\
$^{1}$Space Astronomy Group, ISITE Campus, U. R. Rao Satellite Centre, Outer Ring Road, Marathahalli, Bangalore, 560037, India\\
$^{2}$Dept of Earth \& Space Sciences, Indian Institute of Space Science and Technology, Thiruvananthapuram, 695547, India\\
$^{3}$Albanova University Centre, KTH PAP, Stockholm, 10691, Sweden \\
$^{4}$Department of Astronomy and Astrophysics, Tata Institute of Fundamental Research, Homi Bhabha Road, Colaba, Mumbai, 400005, India\\
$^{5}$Department of Physics, Indian Institute of Science, Bangalore, 560012, India\\
}

\date{Accepted 2019 May 10. Received 2019 May 10; in original form 2019 January 21}

\pubyear{2019}

\begin{document}
\label{firstpage}
\pagerange{\pageref{firstpage}--\pageref{lastpage}}
\maketitle

\begin{abstract}	
We present the results of Target of
Opportunity (ToO) observations made with {\it AstroSat} of the newly discovered black hole binary
MAXI J1535$-$571. 
We detect prominent C-type Quasi-periodic Oscillations (QPOs)
of frequencies varying from $1.85$~Hz to $2.88$~Hz, along with distinct harmonics in all the {\it AstroSat}
observations. We note that while the fundamental QPO is seen in the $3 - 50$~keV energy band, 
the harmonic is not significant above $\sim 35$~keV. 
The {\it AstroSat} observations were made in the hard intermediate
state, as seen from state transitions observed by {\it MAXI} and \textit{Swift}. 
We attempt spectral modelling of the broadband data ($0.7 - 80$~keV) provided by {\it
AstroSat} using phenomenological and physical models. The spectral modelling
using \texttt{nthComp} gives a photon index in the range between $2.18 - 2.37$ and electron temperature
ranging from $21$ to $63$~keV. 
The seed photon temperature is within $0.19$ to $0.29$~keV. 
The high flux in $0.3 - 80$~keV band corresponds to a luminosity varying from $0.7$ to $1.07$~$L_{Edd}$ assuming the source to be at a distance of 8~kpc and hosting a black hole with a mass of $6~M_{\odot}$. 
The physical model based on the two-component accretion flow gives disc accretion rates as high as $\sim 1~\dot{m}_{Edd}$ and
halo rate $\sim 0.2~\dot{m}_{Edd}$ respectively. The near Eddington accretion rate seems to be the main reason for the
unprecedented high flux observed from this source. The two-component spectral fitting of {\it AstroSat} data also provides an
estimate of a black hole mass between $5.14$ to $7.83~M_{\odot}$.

\end{abstract}




\section{Introduction}
X-ray binaries (XRBs) are astrophysical systems that are luminous
in X-rays. An XRB hosts two stars of which one is either an accreting neutron star or black hole 
and the other is the mass donor which is referred to as the secondary star. XRBs are classified into low mass X-ray binaries (LMXBs) or
high mass X-ray binaries (HMXBs) based on the mass of the
secondary star \citep[][and references therein]{Seward2010}.
HMXBs tend to be persistently bright, whereas most of the LMXBs are transient in nature. 
LMXBs can remain in quiescence (below detection limit of the monitoring
instruments) for a very long time and suddenly enter a transient phase
exhibiting an outburst in the flux emitted. This increase in flux 
extends over a period of weeks to months, followed by a slower decay. 
Several hitherto unknown sources have been detected during such outbursts.
Catalogs of black hole binaries discovered till 2016 can be
found in \cite{Corral2016,Tetarenko2016}. Prompt follow up observations, 
in different energy bands are required to understand the nature of the
source and deduce the binary properties.

Emission from black hole binaries can be of thermal and non-thermal origins.
The Keplerian accretion disc \citep{ShaSu1973} around a black hole is considered
to be composed of concentric regions with different temperature, each radiating thermally
and its total spectrum is approximated by a multi-color disk blackbody model \citep{Mitsuda1984}.  
Inverse-Comptonisation \citep{Sunyaev1980, Titarchuk1994, Tanaka1995} of the soft photons from the disc by a hot corona 
contributes to the higher energy thermal or non-thermal emission from the X-ray binary. 
The study of black hole binaries helps us to understand the nature of accretion processes in astrophysics. 
We can also infer the rate of mass accretion from the  disc and corona in the source observed and understand their 
relative contributions to the flux. Spectral and temporal analysis gives astronomers an idea about the state evolution of the source during
an outburst. Besides this, multi-wavelength observations are useful in studying the mechanism of outflows and jet ejection from these sources.

A plot of the hardness ratio versus the total flux is called the hardness intensity diagram (HID) \citep{Homan2001} where hardness is the ratio of flux in hard band to that in soft band.
It is generally observed that the HID of a complete outburst follows a `q' profile
\citep{Homan2005,Nandi2012,Motta2012,Nandi2018,Rad2018}. Using HID and spectro-temporal
parameters, one can classify the outburst periods of the black hole binaries into 
low hard state (LHS), hard intermediate state (HIMS), soft intermediate state (SIMS)
and soft state \citep{Homan2001, Homan2005,2006ARA&A..44...49R, Nandi2012}.

Apart from the HID,
the rms intensity diagram (RID) and the hardness rms diagram (HRD) 
\citep{Motta2012, Belloni2016} can also be used for state classification.
Besides this, the presence of low frequency QPOs (LFQPOs) can indicate the state of
a black hole binary system. C-type QPOs \citep{Casella2005} are usually observed in 
LHS and HIMS whereas the presence of A and/or B-type QPOs correspond to
SIMS \citep[and references therein]{Rad2016, Belloni2016}.

The X-ray transient source MAXI J1535$-$571 was detected independently by
{\it MAXI} (Monitor of All-sky X-ray Image) and {\it Swift/BAT} on Sep~$2$,~$2017$ (\cite{2017ATel10699....1N} and
\cite{2017ATel10700....1K}) at (RA, DEC) = ($15~35~19.73,
-57~13~48.1$).  
\cite{2017ATel10708....1N} suggested that it could be a black hole LMXB, based on the observed rapid random X-ray variability and column
density. The unprecedented large flux levels prompted a flurry of observations made by different observatories as detailed below.
A strong radio counterpart was detected using {\it ATCA} data with spectral \texttt{powerlaw}
index of $0.09$~$\pm$~$0.03$. This suggested emission from a compact radio jet
\citep{2017ATel10711....1R}. {\it Swift/XRT} data in the rising phase of the
outburst revealed C-type QPOs at $4.00 \pm 0.05$~Hz, $3.67 \pm 0.05$~Hz
and $2.50 \pm 0.02$~Hz \citep{2017ATel10899....1R}, and evolution of QPOs from $0.21 \pm 0.03$~Hz
to $3.22 \pm 0.04$~Hz \citep{2017ATel10734....1M, 2018AstL...44..378M} using {\it Swift} and {\it INTEGRAL} data. 
\cite{2018arXiv180607147S} also observed QPOs in the range between $0.44$ to $6.48$~Hz using {\it Swift/XRT} data 
in the period MJD 58004 to 58050.
{\it MAXI} and
{\it Swift} monitoring of the outburst as it reached its peak in about $20$
days since detection showed state changes \citep{2017ATel10729....1N,
2017ATel10731....1K, 2017ATel11020....1S}.
\cite{2018ATel11568....1N} reported
that the source flux dropped sharply by MJD 58200 (by Mar~$23,~2018$)
below detection limit in {\it MAXI}, while the source was in the soft state.
X-ray and Radio detections with state changes were nevertheless
reported beyond this too \citep{cospradio, 2018ATel11611....1R,
Parikh2018, 2018ATel11682....1N}.

\cite{2018arXiv180400800N} provided the details of monitoring with {\it MAXI}
$2 - 20$~keV energy band, and the state changes 
till fading of the source. \cite{2018ApJ...852L..34X}
described the {\it NuSTAR} spectra considering the disc reflection features 
while modelling the broad Fe-K line and Compton hump in $20 - 50$~keV. 
They deduced a black hole spin $a$ > 0.84, inner disc
radius $R_{in} < 2.01~r_{ISCO}$, and a lamp-post height
$h=7.2^{+0.8}_{-2.0}~r_{\rm g}$ for a compact coronal illuminating
source. \cite{Tao2018} made use of the
{\it XRT} data from the entire outburst phase to examine the hardness
based state changes, and evolution of spectral properties. 
\cite{baglio18} discussed about the possibility of
using rapid variability in mid-IR in order to study the disc-jet
connection. Recently, \cite{Stiele2018} have done detailed timing and spectral
analysis of this source using data from {\it XRT}, {\it XMM-Newton} and {\it NICER} observatories.
Also, \cite{Stevens2018} detected LFQPO of $5.72^{+0.04}_{-0.06}$~Hz in the average power spectrum 
corresponding to the {\it NICER} observations from MJD 58016 to 58026, when the source was in SIMS. 
\cite{2018arXiv180805318H} reported on timing and spectral
analysis using data from {\it HXMT} and suggested that MAXI J1535$-$571 could be a high inclination system. Analysis of {\it NICER} spectrum
in the energy range $2.3 - 10$~keV by \cite{Miller2018} also indicated a high inclination of $67.4 \pm 0.8 ^{\circ}$. Besides this, \cite{Miller2018}  determined that the source spins at near maximal rate of $0.994 \pm 0.002$.
Though there are estimates on the inclination and spin of the system, the distance and mass of the black hole are yet to be estimated.
In this work, we have used our physical model to estimate the mass of MAXI J1535$-$571. 

{\it AstroSat} ToO observations were triggered on MAXI J1535$-$571, in the second week
after its detection. {\it AstroSat} is India's first satellite, dedicated to astronomy.
The broadband coverage and high sensitivity of {\it LAXPC} and {\it SXT} onboard {\it AstroSat}, enabled simultaneous spectral and temporal studies of the source. We studied the nature and energy dependence of LFQPOs and the broadband spectrum using these observations. We modelled the broadband spectrum in the energy range $0.7 - 80$~keV phenomenologically as well as in the framework of the two-component accretion flow \citep{1995ApJ...455..623C,2006ApJ...642L..49C,iyer2015determination}. This allowed us to compute the accretion rates for the source and also to estimate the mass of the black hole.
In order to augment the 6 days of {\it AstroSat} observations, we also used data from observations with {\it Swift/XRT} and {\it MAXI}.

In \S \ref{sec:obs}, we have given details on the observation and steps for data reduction.
We have discussed the outburst profile and HID of this outburst in \S \ref{ss:HID_MAXI}. 
Detailed temporal analysis was done with {\it AstroSat/LAXPC} and the results are presented in \S \ref{s:res_time}. 
We undertook a broadband spectral analysis using both {\it SXT} and
{\it LAXPC} data and the results from phenomenological as well as physical modelling are presented in \S \ref{ss:spec_analysis}. Finally, we have discussed the results and have concluded in \S \ref{sec:disco}.

\section{Observation and Data Reduction}
\label{sec:obs}

{\it AstroSat} provides a unique platform for simultaneous observations, over a 
wide X-ray band from $0.3$~keV to $100$~keV, via its suite of co-aligned
X-ray instruments -- Soft X-ray Telescope ({\it SXT}) \citep{Singh2016,Singh2017},  
Large Area X-ray Proportional Counter ({\it LAXPC}) \citep{Yadav2016Proc,Antia2017} and  
Cadmium Zinc Telluride Imager ({\it CZTI}) \citep{Vada2016}. 
We exploit the detection capabilities of {\it SXT} in the soft X-rays to obtain spectra in the lower energies and {\it LAXPC} enables us to study spectra upto 80~keV.
We also make use of the temporal resolution ($10~ \mu$s) of {\it LAXPC} to search for QPOs across a wide range of frequencies.

The ToO observations of {\it AstroSat} were scheduled over $62$ slots, each
of about a few ks exposure (Table~\ref{tab:astrosatobsall}), intermittently between Sep~$12$,
$2017$ (MJD~$58008.23$) till Sep~$17$, $2017$ (MJD~$58013.15$). These
were the days when the source was reported by monitoring observations
to be softening compared to the initial hard state though the spectrum 
below 10~keV remained non-thermal 
\citep{2017ATel10729....1N, 2017ATel10731....1K, 2017ATel10733....1P}. 
We made use of data from {\it SXT} and {\it LAXPC}, which operate respectively
in $0.3$~keV to $8$~keV and $3$~keV to $80$~keV bands, for the timing and spectral analyses. 
The {\it AstroSat} level~1 ToO data were obtained from the ISSDC data dissemination
archive\footnote{https://gads.issdc.gov.in/astro-gads/}. 
Table~\ref{tab:astrosatobsall} lists details of all the {\it AstroSat} observations of MAXI J1535$-$571, including orbit numbers and exposure time.

The {\it Swift} mission, as mentioned earlier, continuously monitored the
source from its first detection of the outburst.
We made use of the {\it XRT} data in the $0.3$~keV to $10$~keV band, 
in order to augment the {\it AstroSat}
data used in the analysis. The {\it XRT} data were obtained from the HEASARC
public archive\footnote{https://heasarc.gsfc.nasa.gov/cgi-bin/W3Browse/w3browse.pl}. 
Details of all publicly available {\it Swift} monitoring data
are provided in Table~\ref{tab:swiftobsall}.

We also used data from {\it MAXI}\footnote{http://maxi.riken.jp/mxondem/} 
to plot the outburst flux profile and the HID. 
Besides these X-ray instruments, we used radio results
from the Australia Telescope Compact Array ({\it ATCA}) and Atacama Large
Millimeter/submillimeter Array ({\it ALMA}) to study the radio flux levels
during the outburst. Note that the radio results used are the ones available in
\cite{2017ATel10711....1R, 2017ATel10899....1R, 2017ATel10745....1T}. There are more observations reported by ALMA \citep{cospradio}.
However, those results are not yet published. 

\subsection{{\it AstroSat} Data Reduction}

Firstly, we carried out temporal and spectral analysis of the
source MAXI J1535$-$571 using {\it AstroSat} data. The capabilities of {\it LAXPC}
for timing and spectroscopic studies of various sources are already
demonstrated by \cite{Yadav2016,Misra2017,Chauhan2017}. We followed these
papers and the instructions\footnote{http://astrosat-ssc.iucaa.in/?q=laxpcData} provided along with the {\it LAXPC} analysis software (\texttt{LaxpcSoft}\footnote{http://www.tifr.res.in/$\sim$astrosat\_laxpc/LaxpcSoft.html}) released on  May 19, 2018.
The details of how the responses and background spectra are generated are given by \cite{Antia2017}.

The three {\it LAXPC} detectors ({\it LAXPC10, LAXPC20} and {\it LAXPC30}) were operated
in the normal or Event Analysis (EA) mode, 
in all those observations between Sep~$12$ -- Sep~$17$, $2017$.
Of the $62$ sets of data in Table~\ref{tab:astrosatobsall},
we used $15$ for our analysis, sampling a few from the several orbits of observation per day. 
Selected orbit numbers have been bold-faced in the table. As will be evident from the 
results in \S \ref{s:res_time} and \S \ref{ss:spec_analysis}, the temporal and spectral parameters do not evolve 
indicating that the selected observations are enough to characterise the source.

Of the three {\it LAXPC} detectors, data from {\it LAXPC10}, and in that, the
eventlist subset corresponding to the longest available good time
interval slot has been used for analysis. The lightcurve has been
extracted in the $3 - 80$~keV energy range with a resolution of
$0.005$~s which corresponds to power spectra ranging upto a frequency
of $100$~Hz. The background subtracted mean count rate for the
{\it LAXPC10} observations varied from around $4400$~cts/s to
$6300$~cts/s. We also generated lightcurves in the energy bands
$3 - 20$~keV, $20 - 35$~keV, $35 - 50$~keV and
$50 - 80$~keV for selected observations to study the energy
dependent evolution of the features in the power spectrum.
Further, the source and background energy spectra were generated.
Out of the three layers of {\it LAXPC}, we used data from the top layer only.
Moreover, only single events were considered for generating the spectra.
These options were selected to have a minimal undesired bump in the spectral residuals at around 33~keV. 
A systematic error of $2\%$ has been added while creating the
background spectrum. \texttt{XSPEC v 12.9.1} has been used for
spectral analysis and modelling.

\begin{figure}
	\includegraphics[width=\columnwidth]{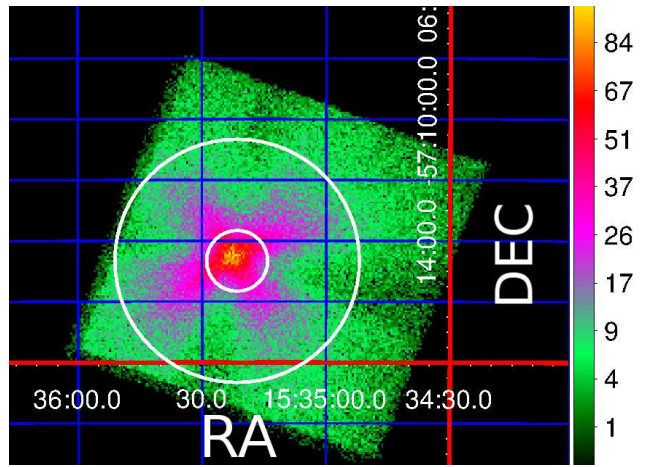}
	\caption{Image of the source MAXI J1535$-$571 taken on MJD 58009.55 using the {\it SXT}
	instrument aboard {\it AstroSat}. In order to address the pile-up issue in the {\it SXT} 
	Fast Windowed mode data, we have selected the source region as an annulus with inner and
	outer radii $1 ^{\prime}$ and $4 ^{\prime}$ respectively, as shown. The colourmap
	alongside is for the source intensity in {\it SXT} cts pixel$^{-1}$.}
	\label{fig:sxt_annulus}

\end{figure}
 

{\it SXT} is an instrument capable of 
X-ray imaging and spectroscopy
in the $0.3 - 8$~keV energy range. {\it SXT} has a focusing telescope
and a CCD detector \citep{Singh2017}. 
Extraction of level~$2$ data from {\it SXT} has been done using the
\texttt{sxtpipeline} tool. The {\it SXT} data are available in the Fast Windowed
(FW) mode. The extracted cleaned event file is used to generate energy
spectra with \texttt{XSELECT V2.4d}. The image count rate at the central
pixels goes as high as 369~cts/s which crosses the pile-up limit
of 344~cts/s for FW mode as indicated in 
the \textit{AstroSat Handbook}\footnote{http://www.iucaa.in/$\sim$astrosat/AstroSat\_handbook.pdf}.
To avoid pile-up effects, an annular region centred at the source RA and DEC,
and of $1 ^{\prime}$ inner radius and $4 ^{\prime}$ outer radius is chosen to extract the
spectrum. In Figure \ref{fig:sxt_annulus}, we have shown an example of choosing source
region excluding the central $1 ^{\prime}$ from {\it SXT} data. It is evident from the figure that
the largest radius for the annulus possible in this mode is $4 ^{\prime}$. The average count rate
of the entire image is 175~cts/s and it reduces to 101~cts/s after filtering the region using the
annulus as mentioned above. We used the background, response
and ARF files provided by the {\it SXT} instrument team\footnote{http://www.tifr.res.in/$\sim$astrosat\_sxt/dataanalysis.html} 
for the analysis.

\begin{figure*}
	\includegraphics[width=\textwidth]{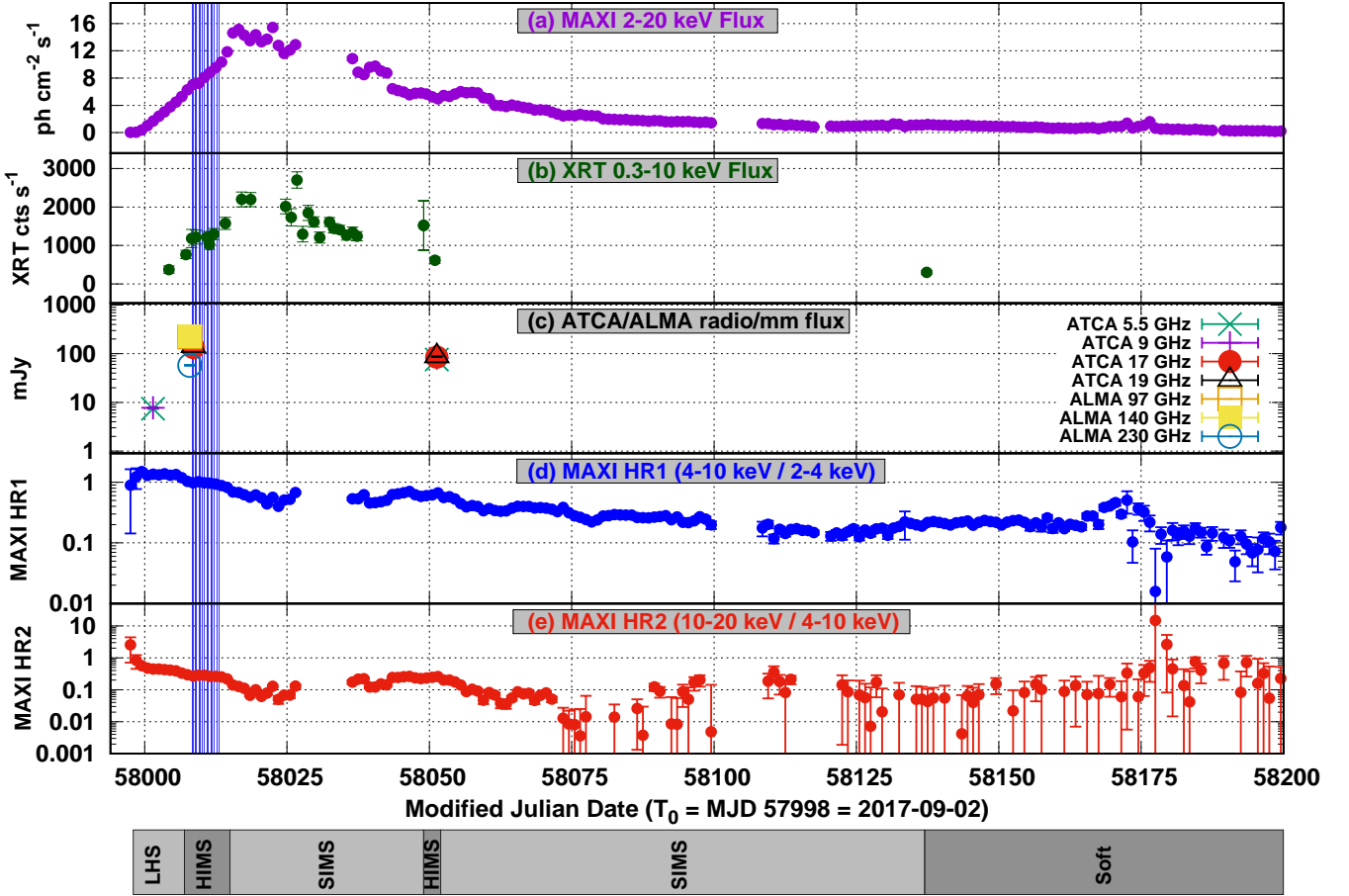}
	\caption{{\it AstroSat} observation slots between Sep~$12$ till Sep~$17$, 2017
		 indicated as vertical blue bars in all panels. (a) {\it MAXI} $2 - 20$~keV light curve
		 of the outburst; (b) {\it XRT} $0.3 - 10$~keV light curve; (c) {\it ATCA/ALMA}
         radio detections of the source (observations reported
         in \citet{cospradio} are not included); (d) {\it MAXI} HR1 ($4 - 10$~keV / $2 - 4$~keV);
		 and (e) {\it MAXI} HR2 ($10 - 20$~keV / $4 - 10$~keV). Indicated at the bottom is the sequence of 
		 different spectral states of the system during the evolution of the outburst, 
		 as determined in this work, and also commensurate with \protect\cite{Tao2018}.}
	\label{fig:fig1}

\end{figure*}

\subsection{{\it Swift/XRT} Data Reduction}
\label{ss:swiftred}

Among all the {\it Swift} observations of the source in the public archive,
the ones that we selected for analysis are indicated by bold-faced ObsIds
in Table~\ref{tab:swiftobsall}.
All {\it XRT} observations from 02 Sep 2017 to 10 Mar 2018 with an exposure greater
than 800~s were considered for this analysis. 
For temporal analysis, lightcurves, power spectra
and hardness ratios of all datasets were obtained. 
{\it XRT} data extraction was done using steps outlined in the {\it XRT} reduction 
manual\footnote{https://swift.gsfc.nasa.gov/analysis/xrt\_swguide\_v1\_2.pdf}. 
\texttt{HEAsoft v 6.21} was used to reduce the data with CALDB updated till Jan 2018. 
All {\it XRT} data used were taken in the Windowed Timing (WT) mode. Initial reduction was done using the 
ftool \texttt{xrtpipeline} with event grades between $0 - 2$ chosen. The cleaned event files so
obtained were used for further analysis.

For generating the power spectra, lightcurves
with a bin size of 0.005~s were made using photons in the energy range $0.5 - 10$~keV 
and $3 - 10$~keV (for compatibility with {\it AstroSat} results). All observations with
an average count rate $>$ 100~cts/s were treated for pile-up effects. 
We selected the source extraction region using an annulus of inner radius $12.5 ^{\prime \prime}$ and outer radius $50 ^{\prime \prime}$, centred on the source position. Data from the region within the inner radius of the annulus was discarded to correct for pile-up. 
The power spectra were generated from these pile-up corrected lightcurves as
explained in \S \ref{ss:flux_rms}.
Lightcurves with a time bin of 1~s in the bands $0.5 - 4$~keV and $4 - 10$~keV 
were created to compute the hardness ratio between these bands. 
In the next section, we present the
outburst profile and HID.


\section{2017 outburst profile and Hardness Intensity}
\label{ss:HID_MAXI}


Figure \ref{fig:fig1} shows the {\it MAXI} outburst profile in $2 - 20$~keV, 
{\it XRT} light curve ($0.3 - 10$~keV), 
radio detections made using {\it ATCA} \& {\it ALMA}, 
{\it MAXI}~ HR1 ($4 - 10$~keV / $2 - 4$~keV) and 
{\it MAXI}~ HR2 ($10 - 20$~keV / $4 - 10$~keV), respectively, from top to bottom.
MJDs corresponding to 
{\it AstroSat} observations are shaded in all the panels. It can be noticed that the
{\it MAXI} and {\it XRT} flux vary similarly. 
Radio flux of
the order of a few hundred mJy was observed just before the observations
by {\it AstroSat}. 
It is to be noted that the radio activity observed in black hole binaries are strongly correlated with spectral states \citep{Rad2016},
and thus the radio results confirm that MAXI J1535$-$571 was making a transition to HIMS in that period. 
HR1 decreased from a high value of around 1.0 at the beginning
of the outburst to below 0.1 as the flux reduced considerably. Meanwhile HR2
remained low throughout the outburst except during the initial few days.

\begin{figure} 
	\includegraphics[width=.45\textwidth]{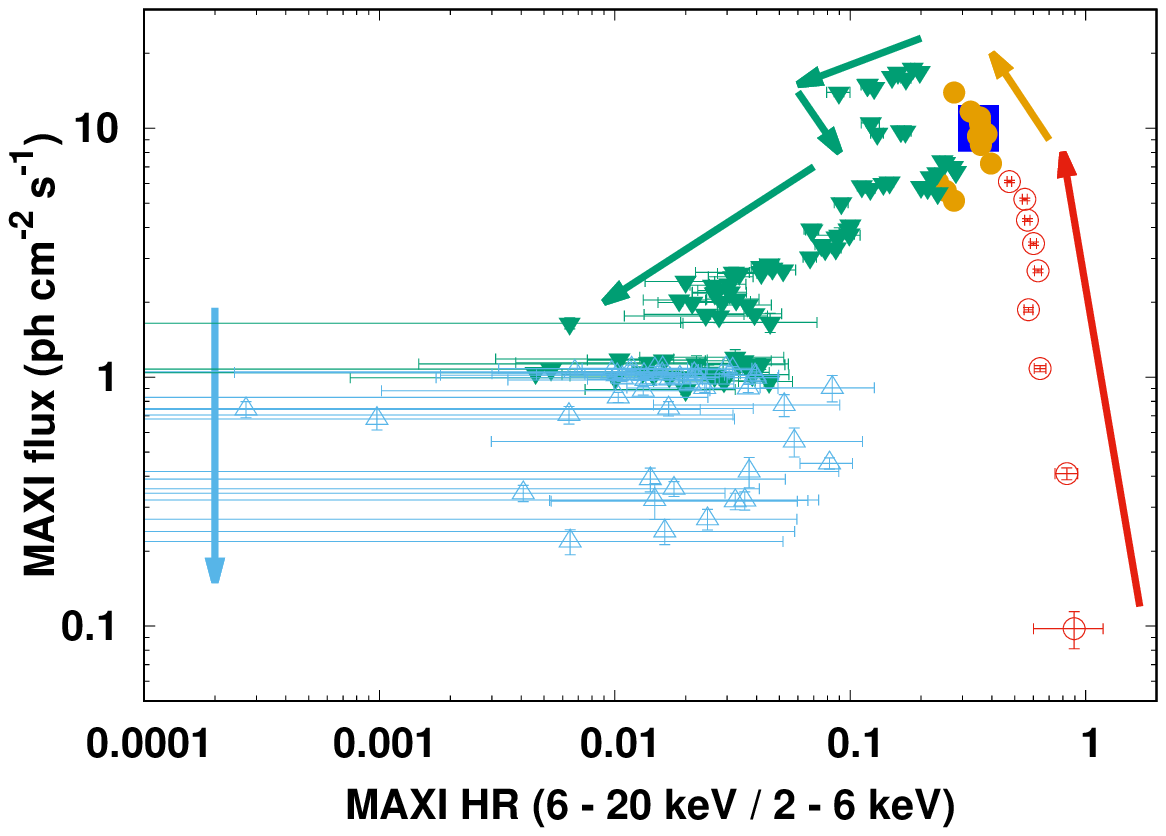} \hfill
	\caption{MAXI Hardness-Intensity Diagram (HID) showing evolution of the hardness
		 ratio ($6 - 20$~keV / $2 - 6$~keV) with the MAXI $2 - 20$~keV Flux. 
		 Here LHS is represented by red circles,
         HIMS by golden filled circles, SIMS by turquoise filled inverted  triangles,
         and soft state is represented using cyan triangles. Times corresponding to 
         {\it AstroSat/LAXPC} observations are shown as a blue patch. 
         The arrows indicate the direction in which the outburst evolves with time.}
	\label{fig:MAXIXRTHID}
\end{figure}
%

Figure \ref{fig:MAXIXRTHID} shows the Hardness Intensity Diagram (HID) 
produced from the same {\it MAXI} data points used in Figure~\ref{fig:fig1}. 
The hardness ratio ($6 - 20$~keV
/ $2 - 6$~keV) is plotted against MAXI $2 - 20$~keV band flux.
The uncertainty on the {\it MAXI} flux increased as the outburst activity dropped
beyond MJD $\sim\:58100$ and that in turn caused the significant spread
in the HID regime of flux $\leq\:2$ ph cm$^{-2}$ s$^{-1}$ and hardness
$\leq\:0.1$. The {\it LAXPC} observations were made just before the outburst peak, 
when the source was in HIMS with a hardness around $0.3$. 
In this figure, LHS is represented by red circles,
HIMS by golden filled circles, SIMS by turquoise filled inverted triangles,
and soft state is represented using cyan triangles. The states are marked
as per the classification done by \cite{Tao2018}. 


The blue shaded region representing {\it AstroSat} observations, falls in HIMS of the outburst.
Beyond the HIMS, the source hardness
decreased as can be seen from Figure~\ref{fig:MAXIXRTHID}. In this phase starting from MJD 58015, the source
entered SIMS \citep{2018arXiv180805318H}. The HID for this
source does not follow the exact `q' - shape that is exhibited by canonical
black hole binaries \citep{belloni2005evolution, Nandi2012, Rad2018}. Instead, we noticed an increase in hardness ratio after
the source reached its peak indicating a transition back to the HIMS. This continued for a few days (MJD 58049 to 58051) before
the hardness ratio started decreasing again. Though there is no XRT observation following this until MJD 58137,
the decrease in hardness ratio from {\it MAXI} HID indicates that the source had gone back to SIMS in
this period.  After that, both flux and hardness ratios decreased considerably.
\cite{Tao2018} have classified this period as soft state (from MJD 58137 to 58230).
In \S \ref{s:res_time} we present the procedure and results of temporal analysis of this outburst.  
 



\section{Timing Analysis and Results}
\label{s:res_time}
\subsection{Timing Analysis}
\label{ss:ta}
Initially, we generated 1 s lightcurves from {\it LAXPC} and obtained the count rates
corresponding to each observation. We did not observe any signature of structured variability in these lightcurves.
Then we created $0.005$~s resolution lightcurves from {\it LAXPC10} data sets to generate the power spectra.
We divided each lightcurve in the intervals of $8192$ time bins, and then
created power density spectra (hereafter PDS) for each interval. The final PDS for each lightcurve
was generated after averaging these individual PDS. 
The averaged PDS was binned by a factor
of $1.05$ in frequency space. We have fitted the PDS with the
combination of $4$~ \texttt{Lorentzians} representing a broad band limited
noise (BLN; a zero centered \texttt{Lorentzian}), a low frequency noise (LFN),
QPO and its upper harmonics. We took the QPO frequency as the centroid of the \texttt{Lorentzian} component fitting the QPO. 
QPOs were identified by fitting a multi-Lorentzian model 
to the entire power spectrum and searching 
for features with \textit{quality factor}, ~$Q=\frac{\nu}{\Delta\nu} > 3$ and $Significance=\frac{Norm}{Err_{norm}} > 3$,
where $\nu$ is \texttt{Lorentzian} centroid frequency, $\Delta \nu$ is FWHM and $Err_{norm}$
is the negative error in norm.

\begin{figure}
\includegraphics[height=\columnwidth, angle=-90]{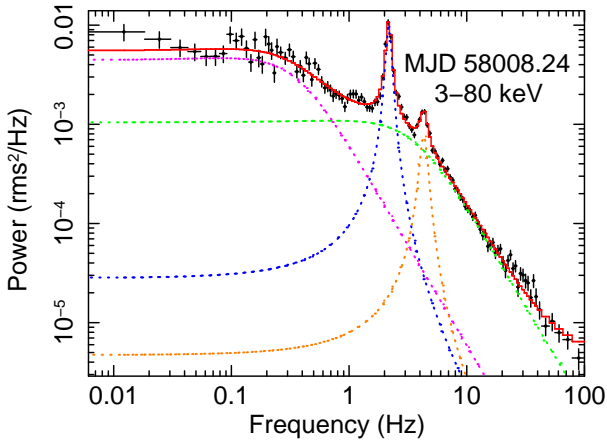}
\caption{Power spectrum of MAXI J1535$-$571 for the energy band $3 - 80$~keV corresponding to the {\it LAXPC}
observation on MJD 58008.24 (Orbit 10585) fitted with multiple \texttt{Lorentzians}. 
A QPO at $2.21 \pm 0.02$ Hz and its harmonic are evident in the PDS. We have also plotted the unfolded
model to show the relative contributions of the model components.}
\label{fig:PDS_full}
\end{figure}

Further, we subtracted the dead-time ($\tau _d$) affected Poisson noise level \citep{2018MNRAS.477.5437A}. 
Besides altering the noise level, the dead-time effects can also change the source rms \citep{VanderKlis1989}. Therefore, we applied correction on rms as prescribed by \cite{Bachetti2015} to account for the effect of dead-time. The dead-time corrected rms ($rms_{in}$) is calculated as $rms_{in} = rms_{det}/(1 - \tau _d \: r_{det})$, where $rms_{det}$ is the detected rms and $r_{det}$ is the detected count rate. We modelled the resultant PDS in frequency-rms space using multiple \texttt{Lorentzians} as mentioned before.
The QPO rms amplitude in percentage was computed by finding the area under the \texttt{Lorentzian} corresponding to the QPO, in the frequency range between 0.05 to 100~Hz.
RMS calculations were done for the
full band ($3 - 80$~keV) as well as for the $3 - 10$~keV using 
{\it LAXPC10} data in the $0.05 - 100$~Hz frequency range. Similarly, we 
computed rms for the {\it XRT} data in the range $0.5 - 10$~keV
as well as for the $3 - 10$~keV band in order to compare with {\it LAXPC10} 
results. Total rms of each PDS was obtained by adding the rms of individual features in quadrature.

\begin{figure*} 
	\includegraphics[trim=0 0 0.25mm 0, clip = true, width=0.45\textwidth, height=6.5cm]{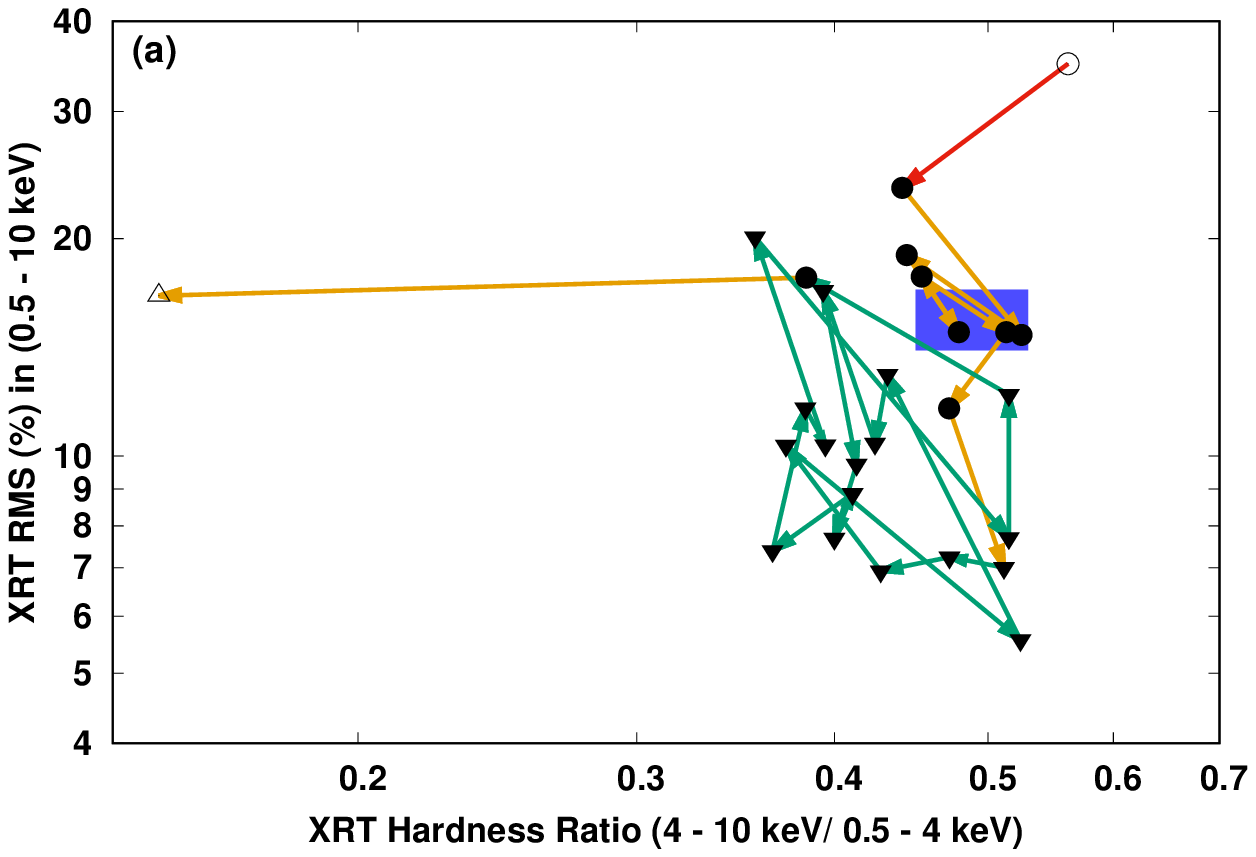}\hfill
	\includegraphics[trim=0 0 0.25mm 0, clip = true, width=0.45\textwidth, height=6.5cm]{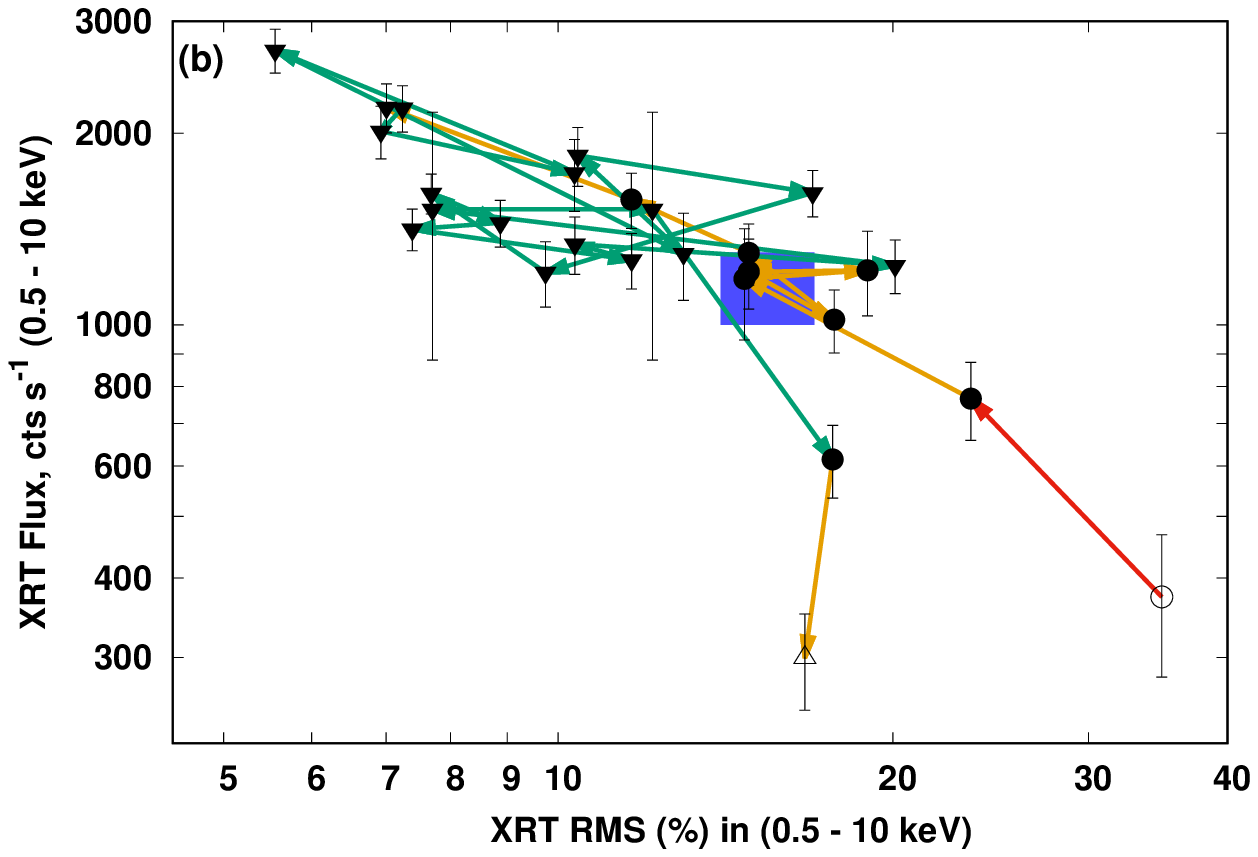}
	\caption{(a) XRT Hardness-RMS Diagram (HRD) and on panel (b) we have the RMS Intensity Diagram (RID).
	Here hollow circles represent LHS, filled circles represent HIMS, inverted triangles stand for SIMS and hollow triangles
	indicate soft state. The arrow colours have same meaning as in Figure \ref{fig:MAXIXRTHID}. 
    Also, times corresponding to {\it LAXPC} observations are shown as blue patches.}
	\label{fig:XRT-HRD-RID}
\end{figure*}


Figure \ref{fig:PDS_full} shows
the power spectrum corresponding to the observation on MJD 58008.24.
This was obtained from a lightcurve in the range $3 - 80$~keV.
The power spectrum in frequency-rms space was modelled with multiple \texttt{Lorentzians} corresponding to
fundamental QPO , its harmonic, BLN and LFN. The fundamental
frequency was found to be $2.21 \pm 0.02$~Hz with an FWHM of $0.24 \pm 0.02$~Hz.
The second harmonic was found at $4.37 \pm 0.03$~Hz with an FWHM of $0.76 \pm 0.10$~Hz.
The percentage rms of the fundamental frequency was $8.23 \pm 0.49$ and for the second harmonic, it was $4.10 \pm 0.37$.
The BLN contributed $11.30 \pm 1.44$\% rms and the LFN had rms of $6.56 \pm 1.02$\%. 
The $\chi ^2/dof$ for this fit was 0.98 (115.7/118).
Unless mentioned explicitly, all the errors are computed using $\Delta \chi ^2 = 1.0$ (68\% confidence level) for all the observations 
(see Tables \ref{tab:qpoparAll} and \ref{tab:qpoparEner}).
The detection of a QPO and its harmonic in this PDS encouraged
us to do an energy dependent analysis of the power spectra, and the details are presented in \S \ref{sec:tempo}.

\subsection{Relation between flux and RMS}
\label{ss:flux_rms}
We generated the power spectra from {\it XRT} lightcurves (pile-up corrected) in the $0.5 - 10$~keV energy range, using the ftool \texttt{powspec}, considering 8192 bins per interval. The dead-time correction factor for {\it XRT} is $\sim$ 1 and so it does not affect the power spectrum. Exposure correction was not applied as it can affect the Poissonian nature of the noise power. 
The power spectrum in frequency-rms space was integrated from 0.05 to 100~Hz to compute the rms value.
The evolution of fractional rms variability of the {\it XRT} observations with respect to hardness and intensity is shown in Figure \ref{fig:XRT-HRD-RID}.
Hardness was calculated as the ratio of flux in $4 - 10$~keV to the flux in $0.5 - 4$~keV. 
Total rms for each PDS was calculated using the \texttt{imodel} command of QDP\footnote{https://heasarc.gsfc.nasa.gov/ftools/others/qdp/qdp.html}.
Here hollow circles represent LHS, filled circles represent HIMS, inverted triangles stand for SIMS and hollow triangles indicate soft state.
The arrows have colours assigned as in Figure \ref{fig:MAXIXRTHID} based on the state at the tail of the arrow.
\cite{Motta2012, Belloni2016} show the general trend of an HRD corresponding to a canonical black hole binary outburst where the 
rms during LHS varies from $\sim 20$ to $40$\%, HIMS from $\sim 10$ to $20$\% and below $\sim 5$\% in the softer states.	  
Figure \ref{fig:XRT-HRD-RID}a shows the HRD for MAXI J1535$-$571 with {\it XRT} data from which it was evident that at the beginning of the outburst in LHS, the hardness (0.56) and rms (around 35\%) are both at a comparatively larger value. In the HIMS, the rms value was observed to be in the range 10 to 20\%. As the outburst progressed, the rms value decreased to around 6\% in SIMS. {\it AstroSat} observations are indicated by the blue shaded portion. These observations were made just before the drop in rms values. 
The observation from soft state shown in the HRD has a significantly high value of rms ($\sim 17$\%) as compared to other sources \citep[see][]{Nandi2012, Motta2012, Rad2018}.  

From Figure \ref{fig:XRT-HRD-RID}b, we can see that in the LHS, though the source had highest rms ($\sim 35$\%), the flux was below $\sim 400$ cts/s.
In HIMS, we observe count rates in between 600 and 1500~cts/s and rms within $\sim 10$ to $25$\%.~RID for other sources \citep{Munoz2011} also have higher count rates in the intermediate states as compared to the hard state. The rms and hardness of the {\it AstroSat} observation period suggests that the source was in HIMS as indicated by the golden arrows and filled circles. This is consistent with the spectral state classification by  \cite{Tao2018}.   
The RID shows dip in rms values in the SIMS. Besides this, the intensity
peaks (above 2000 cts/s) during the time when the rms has lowest values ($\sim 6$\%).

\subsection{Temporal Properties}\label{sec:tempo}

As detailed above, we modelled all the 15 {\it LAXPC} power spectra
with multiple \texttt{Lorentzians} and it was observed that
the frequency of the fundamental QPO varies from $1.85$~Hz to
$2.88$~Hz.
The quality factor (Q) for the QPOs are
around 9.0 and the significance was also high ($\sim$~10.0). The QPO rms amplitude
was near 8.5\% for all the {\it LAXPC} observations. The harmonics have Q-factor around 5.0,
significance around 10.0 and rms varying from 3.68 to 4.85\%. 
Low frequency noise (LFN) with rms around 6.5\%
and band limited noise (BLN) with rms around 10\% are also present in the power spectra. 
These QPO parameters are tabulated in Table
\ref{tab:qpoparAll}. 
There was no indication of increase/decrease of fundamental frequency with time.  
Based on \cite{Casella2005, Belloni2016}, we have classified the QPOs
as C-type as they are in the frequency range from $\sim 2 - 3$~Hz,
have $Q \sim 9$ and rms of around 8.5\%, and appear along with strong band limited flat noise component. 
The strong QPOs observed in this source could be possibly due to the oscillation of post-shock corona 
(hereafter PSC; see \citealt{Das2014}).

\begin{figure*}
	\begin{tabular}{@{}ccc@{}}
	
	    \includegraphics[width=5.8cm]{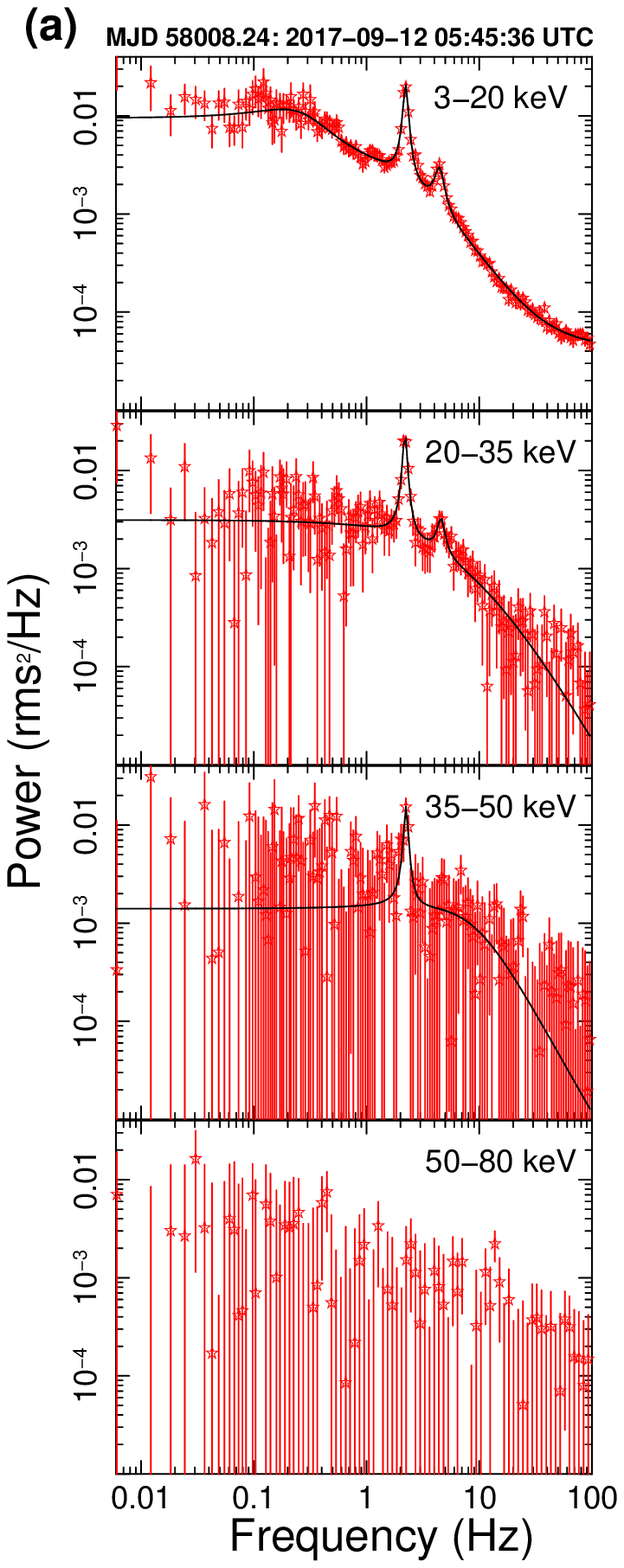}&
	    \includegraphics[width=5.8cm]{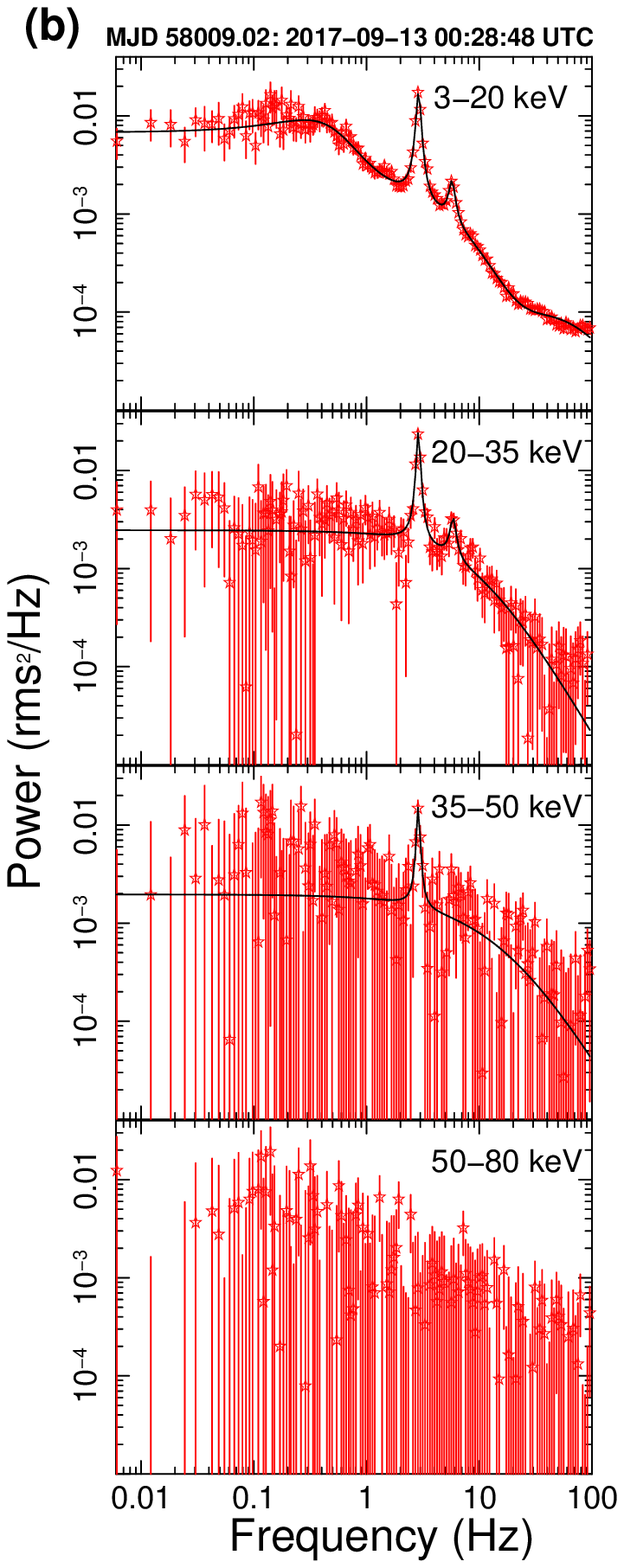}&
	    \includegraphics[width=5.8cm]{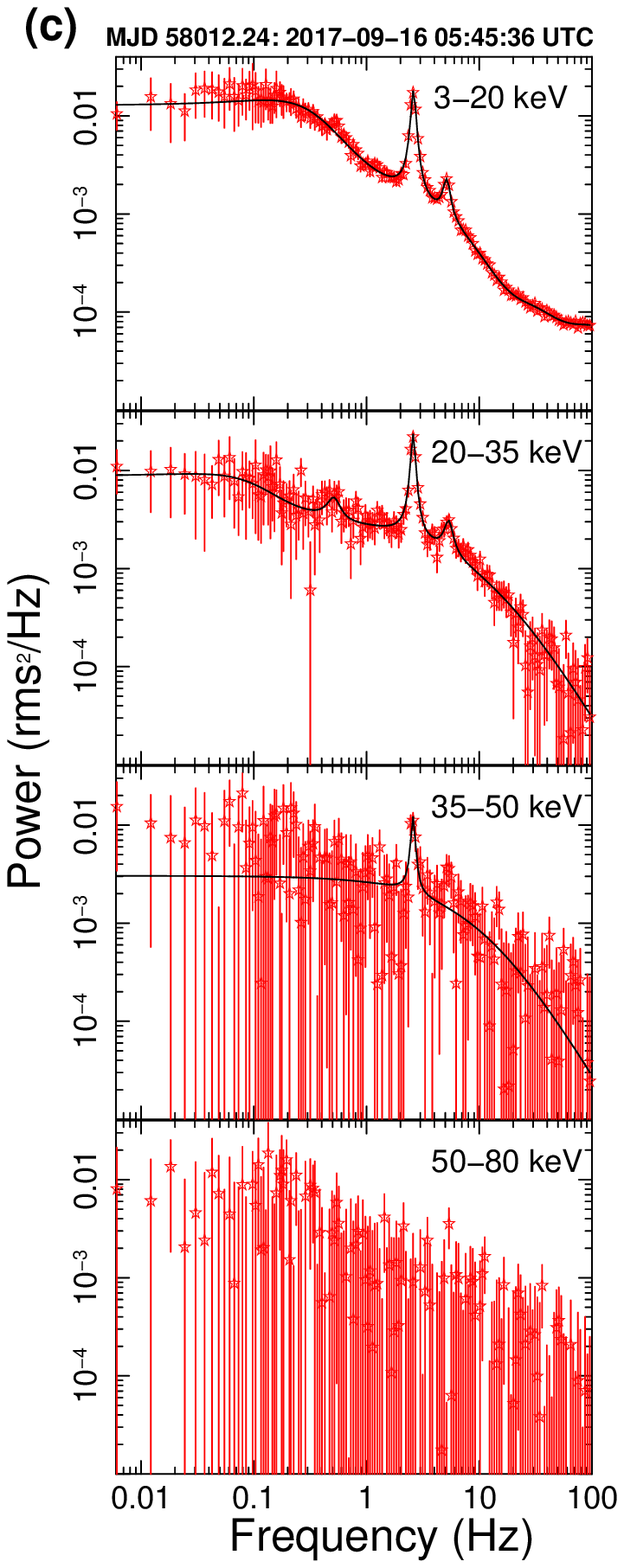}

	\end{tabular}

	\caption{{\it AstroSat/LAXPC} energy dependent PDS: first column for MJD 58008.24 (Orbit 10585)    
	     data, middle for MJD 58009.02 (Orbit 10597) and last column for
	     MJD 58012.24 (Orbit 10644), each for energy bands $3 - 20$~keV,
		 $20 - 35$~keV, $35 - 50$~keV and $50 - 80$~keV
		 respectively. There are no signatures of QPOs
		 in the highest band of $50 - 80$~keV for any of the observed
		 {\it AstroSat} ToO orbits. BLN and LFN are present upto 20~keV. The second harmonic of the  
		 QPO is significant only upto 35~keV. We have not modelled the PDS in $50 - 80$~keV as it is dominated by noise.}
	\label{fig:laxpcpds1}
\end{figure*}

\begin{figure}
	\includegraphics[trim=0 1.2cm 0 0, clip = true,width=1.1\columnwidth]{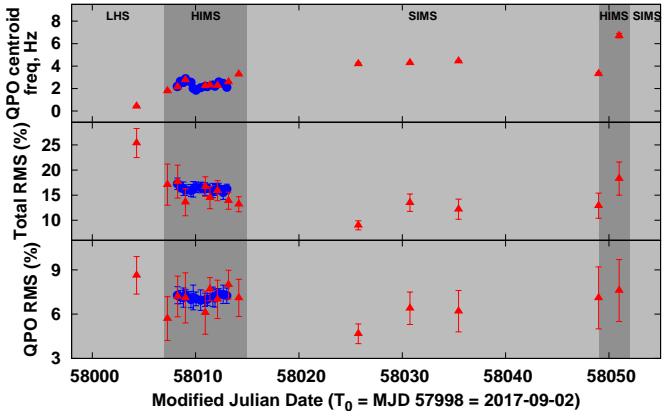}
	\caption{Evolution of QPO frequencies and corresponding RMS
	obtained using {\it LAXPC} (blue) and {\it XRT} (red) $3 - 10$~keV data, 
	are presented in the top and bottom panels respectively.
	The detections with both instruments overlap in the HIMS region.
	{\it LAXPC} results are dead-time corrected and {\it XRT} results are pile-up corrected.  
	The variation of total rms with time is shown in the middle panel. The
	states of the source are also indicated as grey strips in the figure.}
	\label{fig:QPORMSevol}
\end{figure}


  \begin{table*}

	
	\centering
	\caption{QPO parameters for the $15$ orbits data
		 of {\it AstroSat/LAXPC} in the full energy band of
		 $3 - 80$~keV. These results are from modelling the power spectrum 
		 after dead time correction and Poisson noise subtraction.}
	\label{tab:qpoparAll}
	\resizebox{\textwidth}{!}{
	\begin{tabular}{|c|c|c|c|c|c|c|c|c|c|c|c|c|c|}
	\hline

	MJD & Orbit & Mean rate & QPO & Q-factor & Sig$^{b}$ & amplitude & Harmonic & Q-factor & Sig$^{b}$ & amplitude & BLN & LFN & red. $\chi^2$\\
	    &  & (cts/s)& (Hz)&          & ($\sigma$)&  (rms \%) & (Hz)     &          & ($\sigma$)&  (rms \%) &(rms \%) & (rms \%) & ($\chi^2$/dof)\\
	\hline
	58008.24 & 10585 & 4479 & 2.21 $\pm$  0.02  &  9.20 & 11.78  & 8.23 $\pm$ 0.49  & 4.37 $\pm$ 0.03  & 5.75 & 10.30  & 4.10 $\pm$ 0.37 & 11.3 $\pm$ 1.44  &6.56 $\pm$ 1.02  & 0.98 (115.7/118)  \\ 
	58008.53 & 10589 & 5337 & 2.65 $\pm$  0.01  &  8.54 & 13.76  & 8.51 $\pm$ 0.41  & 5.24 $\pm$ 0.04  & 5.88 & 10.63  & 3.68 $\pm$ 0.29 & 10.7 $\pm$ 1.25  &7.03 $\pm$ 0.69  & 0.99 (116.7/118)  \\ 
	58008.82 & 10594 & 5398 & 2.54 $\pm$  0.01  &  8.75 & 09.22  & 8.42 $\pm$ 0.63  & 4.95 $\pm$ 0.05  & 4.54 & 07.75  & 4.32 $\pm$ 0.63 & 9.83 $\pm$ 2.03  &7.37 $\pm$ 0.96  & 0.89 (105.1/118)  \\
	58009.02 & 10597 & 5618 & 2.88 $\pm$  0.01  & 10.28 & 14.10  & 8.56 $\pm$ 0.42  & 5.68 $\pm$ 0.04  & 4.89 & 10.69  & 4.00 $\pm$ 0.39 & 9.91 $\pm$ 1.22  &7.29 $\pm$ 0.64  & 1.39 (164.0/118)  \\
	58009.55 & 10604 & 5593 & 2.56 $\pm$  0.01  &  9.48 & 09.89  & 8.46 $\pm$ 0.66  & 5.02 $\pm$ 0.04  & 4.29 & 08.38  & 4.28 $\pm$ 0.53 & 10.0 $\pm$ 1.63  &6.86 $\pm$ 1.03  & 1.70 (200.3/118)  \\
	58009.77 & 10608 & 5590 & 2.02 $\pm$  0.02  &  8.41 & 07.58  & 8.44 $\pm$ 0.80  & 3.99 $\pm$ 0.06  & 3.21 & 07.30  & 4.85 $\pm$ 0.90 & 9.99 $\pm$ 2.24  &6.50 $\pm$ 1.33  & 0.93 (110.0/118)  \\
	58010.03 & 10612 & 5642 & 1.85 $\pm$  0.01  &  9.73 & 12.53  & 8.22 $\pm$ 0.46  & 3.72 $\pm$ 0.03  & 4.76 & 10.92  & 4.53 $\pm$ 0.39 & 11.2 $\pm$ 1.76  &6.14 $\pm$ 1.48  & 0.83 ( 98.0/118)  \\
	58010.49 & 10618 & 5708 & 2.08 $\pm$  0.01  & 11.55 & 09.87  & 8.42 $\pm$ 0.64  & 4.15 $\pm$ 0.03  & 5.60 & 10.09  & 4.39 $\pm$ 0.40 & 11.2 $\pm$ 1.35  &6.23 $\pm$ 1.08  & 1.09 (128.6/118)  \\
	58010.92 & 10625 & 5910 & 2.20 $\pm$  0.01  &  9.56 & 11.38  & 8.35 $\pm$ 0.42  & 4.33 $\pm$ 0.03  & 5.85 & 09.08  & 4.17 $\pm$ 0.46 & 10.8 $\pm$ 1.61  &6.61 $\pm$ 1.17  & 0.91 (107.3/118)  \\
	58011.13 & 10628 & 5923 & 2.17 $\pm$  0.01  &  9.43 & 13.71  & 8.44 $\pm$ 0.48  & 4.30 $\pm$ 0.02  & 7.16 & 11.80  & 3.87 $\pm$ 0.28 & 11.1 $\pm$ 1.08  &6.61 $\pm$ 0.72  & 1.15 (135.2/118)  \\
	58011.66 & 10636 & 6011 & 2.31 $\pm$  0.01  &  9.62 & 08.61  & 8.73 $\pm$ 0.77  & 4.53 $\pm$ 0.03  & 7.07 & 07.27  & 3.85 $\pm$ 0.48 & 10.4 $\pm$ 2.12  &6.80 $\pm$ 1.31  & 0.68 ( 80.8/118)  \\
	58011.87 & 10639 & 6035 & 2.21 $\pm$  0.01  &  9.60 & 09.85  & 8.62 $\pm$ 0.56  & 4.38 $\pm$ 0.03  & 6.53 & 08.39  & 3.95 $\pm$ 0.45 & 10.5 $\pm$ 1.76  &6.87 $\pm$ 1.12  & 0.78 ( 92.3/118)  \\
	58012.24 & 10644 & 6189 & 2.59 $\pm$  0.01  &  8.63 & 15.52  & 8.86 $\pm$ 0.42  & 5.09 $\pm$ 0.03  & 5.19 & 13.88  & 3.94 $\pm$ 0.31 & 9.93 $\pm$ 1.53  &8.03 $\pm$ 0.73  & 1.66 (197.6/119)  \\
	58012.67 & 10651 & 6282 & 2.46 $\pm$  0.02  &  9.11 & 08.88  & 8.82 $\pm$ 0.69  & 4.83 $\pm$ 0.04  & 5.81 & 07.51  & 3.98 $\pm$ 0.49 & 9.56 $\pm$ 2.25  &7.21 $\pm$ 1.26  & 0.69 ( 81.0/118)  \\
	58013.02 & 10656 & 6338 & 2.11 $\pm$  0.01  &  9.59 & 10.04  & 8.89 $\pm$ 0.58  & 4.20 $\pm$ 0.03  & 4.82 & 11.83  & 4.46 $\pm$ 0.49 & 10.1 $\pm$ 2.34  &7.45 $\pm$ 1.17  & 1.35 (159.4/118)  \\

	\hline

	\end{tabular}
	}

	\footnotesize{
	  \begin{flushleft}
	    $^b$ Significance
	  \end{flushleft}
	}

	
  \end{table*}



  \begin{table*}

	
	\centering
	\caption{Energy dependent QPO parameters of MAXI J1535$-$571 from {\it AstroSat/LAXPC}. Power spectra have been modelled in frequency-rms space. The parameters corresponding to the fundamental component are
		 indicated with (f) and those corresponding to the
		 harmonic component with (h).}
	\label{tab:qpoparEner}
	\resizebox{\textwidth}{!}{
	\begin{tabular}{|c|c|c|c|c|c|c|c|c|c|c|c|c|c|c|c|c|c|c|}
	\hline

	MJD &	E-band & QPO(f) & FWHM(f) & Norm(f) & Q(f) & sig(f)$^a$ & rms(f) & QPO(h) & FWHM(h) & Norm(h) & Q(h) & sig(h)$^a$ & rms(h)\\
		&   keV    & Hz     & Hz      &         &      &  ($\sigma$)&  (\%)  & Hz     & Hz      &         &      & ($\sigma$) & (\%)  \\
	\hline \hline
	    	& 3 -- 20  & 2.22 $\pm$ 0.01 & 0.23 $\pm$ 0.02 & 0.018 &  9.64 & 11.05 & 8.09 $\pm$ 0.46 &4.43 $\pm$ 0.03 &  0.60 $\pm$ 0.08 & 0.0018 & 7.38  & 9.84& 5.09 $\pm$ 0.42 \\
58008.24	& 20 -- 35 & 2.19 $\pm$ 0.01 & 0.21 $\pm$ 0.03 & 0.021 & 10.42 & 7.95 & 8.44 $\pm$ 0.82 &$4.57^{+0.14}_{-0.10}$ & $0.78^{+0.62}_{-0.34}$ & 0.0018 & 5.85  &  3.53& 4.71 $\pm$ 1.62 \\
	    	& 35 -- 50 & $2.22^{+0.07}_{-0.05}$ & $0.31^{+0.12}_{-0.16}$ &  0.010 &  7.16 &  4.47 & 7.12 $\pm$ 2.53 &   -  &  -   &  -   &   -   &    -  &\\  
	    	& 50 -- 80 &   -  &   -  &    -  &   -   &    -  &   -  &  -   &  -   &   -   &   -  & - &  \\
    \hline 
        	& 3 -- 20  & 2.89 $\pm$ 0.01 & 0.29 $\pm$ 0.02 & 0.016 &  9.96 & 11.78 & 8.61 $\pm$ 0.45 &5.72 $\pm$ 0.03 & $0.87^{+0.11}_{-0.09}$ & 0.0013 & 6.57  & 12.76 & 4.17 $\pm$ 0.36\\
58009.02	& 20 -- 35 & 2.88 $\pm$ 0.01 & 0.22 $\pm$ 0.02 &  0.023 & 13.09 & 9.91 & 9.09 $\pm$ 0.69 &$5.86^{+0.10}_{-0.09}$ & $0.82^{+0.50}_{-0.32}$ & 0.0019 & 7.14  &  4.08 & 5.02 $\pm$ 1.43\\
 			& 35 -- 50 & 2.87 $\pm$ 0.03 & $0.19^{+0.06}_{-0.09}$ &  0.014 & 15.10 &  4.28 & 6.61 $\pm$ 2.20 &  -  &  -   &  -   &   -   &    - & \\  
			& 50 -- 80 &   -  &   -  &    -  &   -   &    -  &   -  &  -   &  -   &   -   &   -  & - & \\
	\hline
			& 3 -- 20  & 2.61 $\pm$ 0.01 & 0.28 $\pm$ 0.01 & 0.016 &  9.32 & 15.18 & 8.58 $\pm$ 0.35 & 5.14 $\pm$ 0.02 & $0.84^{+0.11}_{-0.09}$ & 0.0013 & 6.11  & 13.82 & 4.19 $\pm$ 0.29\\
58012.24	& 20 -- 35 & 2.59 $\pm$ 0.01 & 0.26 $\pm$ 0.02 & 0.020 & 9.96 & 11.30 & 9.27 $\pm$ 0.57 & 5.35 $\pm$ 0.10 & $1.09^{+0.45}_{-0.30}$ & 0.0016  &  4.90 & 5.16 & 5.24 $\pm$ 1.05\\
			& 35 -- 50 & $2.58^{+0.04}_{-0.06}$ & $0.24^{+0.07}_{-0.21}$ &  0.0099 &  10.75 &  4.75 & 6.14 $\pm$ 1.93 &   -  &  -   &  -   &   -   &    - & \\  
			& 50 -- 80 &   -  &   -  &    -  &   -   &    -  &   -  &  -   &  -   &   -   &   -  & -& \\
	\hline

	\end{tabular}}

	\footnotesize{
	  \begin{flushleft}
	    $^a$ Significance
	  \end{flushleft}
	}

	
  \end{table*}

Figure~\ref{fig:laxpcpds1} consists of the energy dependent Power Density
Spectra (PDS) obtained with {\it AstroSat/LAXPC} for the following energy
ranges: $3 - 20$~keV, $20 - 35$~keV, $35 - 50$~keV and $50 - 80$~keV,
for MJD 58008.24 (Orbit 10585), 58009.02 (Orbit 10597) and 58012.24 (Orbit 10644). 
Table~\ref{tab:qpoparEner} lists the energy
dependent QPO parameter variations for MJDs 58008.24, 58009.02 and 58012.24. 
It is evident that the fundamental QPO appears in the range
$3 - 50$~keV, while the second harmonic was significant only in the
$3 - 35$~keV band. No QPOs were seen in $50 - 80$~keV band. A low frequency noise (LFN)
and a band limited noise (BLN) also contributed to the power in energy range $3 - 20$~keV.

The evolution of QPO centroid frequency, total rms and QPO rms are presented in Figure \ref{fig:QPORMSevol}.
Here, {\it XRT} results are shown in red while {\it LAXPC} results are shown in blue. In order to compare results, QPO parameters were calculated in 
$3 - 10$~keV for both {\it XRT} and {\it LAXPC}. 
Note that, the {\it XRT} PDS in the $0.5 - 3$~keV band had no QPO like features and these photons piled-up to bring down the QPO rms values in $3 - 10$~keV band. Therefore, we applied pile-up correction as mentioned in \S \ref{ss:swiftred} in order to estimate the correct rms in $3 - 10$~keV {\it XRT} data.
It was observed that the QPO frequency did not evolve except for the first four {\it Swift} observations.
The first QPO of 0.44~Hz was detected with {\it XRT} data on MJD 58004.29 when the source was in the LHS. 
Beyond that, the fundamental frequency was almost constant during the HIMS as indicated by both {\it XRT} and {\it LAXPC} observations.
The total rms for the {\it LAXPC} observations was around $\sim$ $16.2 \pm 1.1$\% and the corresponding QPO rms
remained almost constant at $7.1 \pm 0.52$\%. The total rms in {\it XRT} PDS during the HIMS was around $15 \pm 2.4$\%, 
and the QPO rms was found to have an average value of $6.9 \pm 1.3$\%. The larger uncertainties on width and normalisation of the \texttt{Lorentzians} resulted in larger errors on rms values. This is due to the lower photon statistics of {\it XRT} as compared to the high {\it LAXPC} count rate in $3 - 10$~keV band.  
Only three QPOs ($\sim 4.3$ Hz) were detected between MJD 58014 and MJD 58040 with {\it XRT} data. During this period the source was in SIMS \citep{Tao2018} with very high fluxes (see Figure \ref{fig:fig1}). 
The total integrated rms in $3 - 10$~keV was larger during LHS and HIMS of the outburst than during the SIMS. 
As the source underwent a transition back to HIMS during MJD 58049 to 58051, the rms observed then was around 18\%. We also detected the maximum QPO frequency ($\sim 6.72$~Hz) from the source on MJD 58050.98. QPOs were not observed beyond this date upto MJD 58200. 
The presence of C-type QPOs and their frequencies indicated that the source was likely to be in HIMS during {\it AstroSat} observations. 
In \S \ref{ss:spec_analysis}, we have presented the results of spectral analysis and modelling with the {\it AstroSat} data. 
Further, we intended to estimate the accretion parameters of the source and constrain the mass of the black hole. 
It was with this motivation that we decided to use the two-component flow model (see \S \ref{sec:tcaf}).


%

\section{Spectral Analysis and Results}
\label{ss:spec_analysis}

\subsection{Spectral Analysis}
\label{subs:spectral_analysis}
Initially, we modelled the {\it SXT} spectrum of MJD 58008.24 (Orbit 10585) with \texttt{Tbabs} \citep{Wilms2000}, \texttt{diskbb} and \texttt{powerlaw}. This gave an inner disc temperature of 0.23~keV and a \texttt{powerlaw} photon index
of 2.31 with a $\chi ^2/dof$ of $782.4/612=1.27$.
Energy spectrum from {\it SXT} in the range $0.7 - 7$~keV was then modelled 
with thermally comptonized continuum - 
\texttt{nthComp} (\cite{1996MNRAS.283..193Z},
\cite{1999MNRAS.309..561Z}). 
We used gain correction with a fixed slope of 1 using the \texttt{gain fit}
command to flatten the residuals at $1.8$ and $2.2$~keV. Even
after that, residuals were slightly on the higher side at around
$1.7$ keV and below $0.8$~keV. 
The modelling with \texttt{nthComp}
gave a $\chi ^2/dof$ of $866/613=1.41$.

Then we modelled the corresponding {\it LAXPC10} spectrum 
in the range $3 - 80$~keV initially with \texttt{diskbb + powerlaw} and then with \texttt{nthComp}.
The fit with \texttt{Tbabs}~$\times$~\texttt{(diskbb + powerlaw)} gave poor results at a $\chi ^2_{red} \sim 6$ and it was found that the \texttt{diskbb} component was
insensitive to the fit. 
The fit with \texttt{Tbabs}~$\times$~\texttt{(nthComp)} was better and gave $\chi ^2_{red}$ of $504/274=1.84$. Further, the inclusion of a \texttt{gaussian} to account for
the iron line at around 6.6~keV and an instrumental Xenon \texttt{edge} around 33~keV improved the fit resulting in a $\chi ^2_{red}$ of $272.8/270=1.01$. 
The strong iron line indicated the possible presence of a continuum reflection component in the spectra. In order to check this, we introduced a reflection component (\texttt{ireflect}) and found that the relative reflection coefficient was too low ($0.03 - 0.08$). Moreover, the $\chi ^2/dof = 272.61/269 = 1.01$, did not improve after including reflection component. Therefore, we inferred that a reflection continuum was not required to model the energy spectra of the source using data from {\it AstroSat} observations. The fit results implied that when we combine {\it SXT} and {\it LAXPC} data, it was better to model with \texttt{nthComp} than using the phenomenological model containing \texttt{diskbb} and \texttt{powerlaw}.

A 2\% systematic error in each bin was incorporated in the spectral response as specified in the {\it SXT} user manual\footnote{http://www.tifr.res.in/$\sim$astrosat\_sxt/dataanalysis.html} and by \cite{Leahy2019}. The systematic error for spectral data from {\it LAXPC} detectors was also reported to be within 2\%  \citep{Antia2017}. Therefore, the combined spectral fit with both {\it SXT} and {\it LAXPC} were considered to have a 2\% systematic error in order to achieve a good estimate of the spectral fit parameters.
Once good independent fits were
obtained, we attempted for combined fitting
of energy spectra with {\it SXT} and {\it LAXPC} data using \texttt{nthComp}. The \texttt{gaussian} line width was frozen to 0.6~keV
for the broadband modelling. This gave a reasonable
$\chi ^2_{red}$ of $1181/895=1.31$ with a photon index of $2.254_{-0.005}^{+0.004}$, electron temperature
of $37.8_{-2.37}^{+2.74}$~keV and seed photon temperature of $0.21_{-0.04}^{+0.02}$~keV on MJD 58008.24. 
Unless mentioned explicitly, all the errors are computed using $\Delta \chi ^2 = 1.0$ (68\% confidence level) for all the observations (see Table \ref{tab:specpar}).
Further, we carried out spectral analysis for all the
observations under consideration with \texttt{nthComp} and the results are presented in
\S \ref{sec:spec}.

The hydrogen column density used for this source so far by other investigators 
for spectral fitting are $(4.44 \pm 1.1) \times 10^{22}$ atoms cm$^{-2}$ \citep{2018ATel11611....1R, Parikh2018} and
$4.05 \times 10^{22}$ atoms cm$^{-2}$ \citep{Miller2018}. \cite{2018ATel11611....1R} arrived at this value for $N_H$ (with 25\% error) using three observations from the declining phase of the outburst, when the source was undergoing soft to hard state transition. \cite{Tao2018,2018arXiv180607147S} have left $N_H$ parameter free. It should be noted that the HEASARC $N_H$ calculator\footnote{https://heasarc.gsfc.nasa.gov/cgi-bin/Tools/w3nh/w3nh.pl} gives hydrogen column density of only $1.46 \times 10^{22}$ atoms cm$^{-2}$. The $N_H$ value can change with states of the source due to variation in radio activity (jets/outflows).
As {\it AstroSat} observations of the source were made when the radio flux varied, 
we attempted to determine $N_H$ density independently from the {\it AstroSat} spectra.
The combined fits with {\it SXT} and {\it LAXPC} using free $N_H$ parameter
resulted in $N_H$ values within a narrow range of $2.52 \times 10^{22}$ to $2.67 \times 10^{22}$ atoms cm$^{-2}$.  
As our spectral modelling indicated minimal scatter of the $N_H$ value, the \texttt{Tbabs}
component was left free throughout our analysis. 

\subsection{Spectral Properties}\label{sec:spec}

\begin{figure}
	\includegraphics[height=\columnwidth,angle=-90]{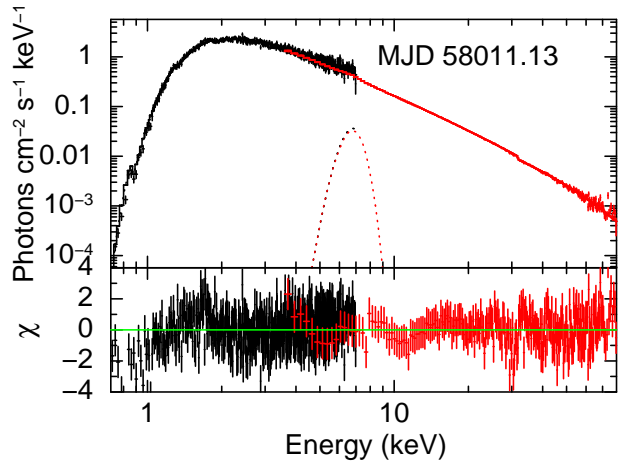}
	\caption{Broadband spectrum in the energy range $0.7 - 80$~keV using data from both
	{\it SXT} ($0.7 - 7$~keV) and {\it LAXPC} ($3 - 80$~keV) onboard 
	{\it AstroSat} corresponding to MJD 58011.13 (Orbit 10628). {\it SXT} data are shown in black and {\it LAXPC} data
	are shown in red. The spectrum is fitted with continuum
	thermal Comptonisation model (\texttt{nthComp}). The source was in HIMS during this observation with a
	photon index of $2.238 \pm 0.005$.}
	\label{fig:broadspec}
\end{figure}

As detailed in \S \ref{subs:spectral_analysis}, the appropriate
model to fit the energy spectra (in the energy range $0.7 - 80$~keV) is \texttt{Tbabs} $\times$ \texttt{constant(gaussian} $+$ \texttt{edge} $\times$ \texttt{nthComp)}. 
In Figure \ref{fig:broadspec},
we have shown the broadband spectrum from MJD 58011.13 fitted with the above model based on \texttt{nthComp}. 
The photon index for this fit is $2.238 \pm 0.005$ with a $\chi ^2_{red}=1.19~(1071/895)$
indicating that the source was in HIMS. 
The results of spectral
modelling and the $\chi ^2 _{red}$ values for all observations are included in
Table \ref{tab:specpar}. 
The entries indicated with a $\dagger$ symbol
correspond to broadband observations combining {\it SXT} and {\it LAXPC} data.
All other entries in this table correspond to {\it LAXPC} observations
alone. The model based on \texttt{nthComp} satisfactorily fits all these
observations. 
We find that the photon index ($\Gamma$) values vary from
$2.18$ to $2.37$ only and did not show much evolution within the
six days of {\it AstroSat} observation. The value of electron temperature, $kT_e$ is in the range between $21.2 - 63$~keV
indicating a minimum high energy cut-off of $\sim 60$~keV ($3~kT_e$). For a few observations, we were unable to constrain $kT_e$
from the fit and in those cases we fixed it to a high value of 100 keV (see Table \ref{tab:specpar}).  
The seed photon temperature parameter $kT_{bb}$, which specifies the low energy roll over varied between $0.19$ to $0.29$~keV.
The norm remained in the range 21.6 to 53.9. A Xenon \texttt{edge} in the range $32 - 34$~keV was required in all cases. 
The \texttt{gaussian} line energy around 6.6 keV had width fixed to 0.6 keV.
The total flux in the $0.3 - 80$~keV during the {\it AstroSat} observations was high and it varied from 
$7.14 \times 10^{-8}$ to $1.10 \times 10^{-7}$~erg~cm$^{-2}$~s$^{-1}$. Assuming the source to be at 8~kpc distance 
as considered by \cite{Tao2018},
we computed the luminosity to be in the range between $5.46 ~\times~ 10^{38}$ to $8.38~ \times ~10^{38}$~erg~s$^{-1}$. 
For a black hole of mass 6~$M_{\odot}$, the luminosity when expressed in terms of Eddington Luminosity is $0.70$ to $1.07$~$L_{Edd}$.
Using the $\Gamma$ and $kT_e$ parameters, we also
computed the optical depth ($\tau$) of the system based on the formula given by \cite{1996MNRAS.283..193Z}. 
$\tau$ was found to be in the range $0.72 - 2.5$. 

As mentioned in
\S \ref{sec:tempo}, the presence of strong QPOs in the power spectrum
helps us to conclude that the corona is oscillating in nature. The lack of
evolution in the QPO frequency and the tight range of spectral photon indices 
both point to the fact that the corona is of almost constant size
during the {\it AstroSat} observations. In the next section, we present the details of spectral 
modelling with two-component accretion flow and determine the
shock location which indicates the size of the PSC. We also attempt to
constrain the mass of the source using the same model.

\begin{table*}

	\centering
	\caption{Spectral parameters resulting from a fit with the model
	 \texttt{Tbabs} $\times$ \texttt{constant} $\times$ \texttt{(gaussian} $+$ \texttt{edge} $\times$ \texttt{nthComp)}. $\dagger$ denotes broadband modelling with {\it SXT} and {\it LAXPC}, whereas other observations are modelled using data
	from {\it LAXPC} instrument alone. Unabsorbed flux values in $0.3 - 80$~keV are also included in the Table.}
	\label{tab:specpar}

    \resizebox{\textwidth}{!}{
	\begin{tabular}{|l|c|c|c|c|c|c|c|l|}
	\hline
		MJD & Orbit & $\Gamma$ & kT$_e$ (keV)& kT$_{bb}$ (keV) & norm & flux ($0.3 - 80$~keV)     & $\tau $ & $\chi^2 _{red}$ \\
   	        &       &          &             &                 &      &(erg~cm$^{-2}$~s$^{-1}$)  &         & ($\chi^2$/dof) \\
	\hline \hline
	
	58008.24 $\dagger$ & 10585  & $2.254_{-0.005}^{+0.004}$ & $37.8_{-2.3}^{+2.7}$    & $0.21_{-0.04}^{+0.02}$   & $34.3_{-1.5}^{+2.6}$    &$7.14 \times 10^{-8}$ & 1.640 $\pm$ 0.011   & 1.31 (1181/895) \\
	58008.53 $\dagger$ & 10589  & $2.346_{-0.005}^{+0.004}$ & 100.0  (fixed)          & $0.19_{-0.04}^{+0.03}$   & $43.5_{-0.3}^{+0.3}$    &$8.68 \times 10^{-8}$ & 0.720 $\pm$ 0.004   & 1.25 (1127/896) \\
	58008.82 $\dagger$ & 10594  & $2.316_{-0.003}^{+0.003}$ & 100.0  (fixed)          & $0.19_{-0.05}^{+0.05}$   & $35.3_{-0.4}^{+0.3}$    &$8.63 \times 10^{-8}$ & 0.739 $\pm$ 0.004   & 1.27 (1140/896) \\
	58009.02           & 10597  & $2.374_{-0.005}^{+0.006}$ & $63.0_{-6.9}^{+10.2}$   & $0.20_{-0.009}^{+0.008}$ & $34.7_{-8.3}^{+2.2}$    &$7.17 \times 10^{-8}$ & 1.010 $\pm$ 0.008   & 1.14 (311/271)  \\ 
	58009.55 $\dagger$ & 10604  & $2.327_{-0.003}^{+0.003}$ & 100.0  (fixed)          & $0.22_{-0.03}^{+0.01}$   & $43.4_{-5.5}^{+5.2}$    &$9.21 \times 10^{-8}$ & 0.732 $\pm$ 0.004   & 1.41 (1268/896) \\
	58009.77           & 10608  & $2.243_{-0.004}^{+0.005}$ & $39.1_{-2.3}^{+2.9}$    & $0.24_{-0.008}^{+0.006}$ & $25.2_{-8.3}^{+4.1}$    &$7.29 \times 10^{-8}$ & 1.612 $\pm$ 0.009   & 1.15 (312/271)  \\
	58010.03           & 10612  & $2.189_{-0.004}^{+0.005}$ & $24.2_{-0.6}^{+0.8}$    & $0.29_{-0.12}^{+0.03}$   & $21.6_{-7.1}^{+0.5}$    &$7.41 \times 10^{-8}$ & 2.380 $\pm$ 0.010   & 1.09 (296/271)  \\
	58010.49 $\dagger$ & 10618  & $2.233_{-0.002}^{+0.005}$ & $32.6_{-1.2}^{+1.3}$    & $0.21_{-0.07}^{+0.01}$   & $38.2_{-0.3}^{+0.3}$    &$9.39 \times 10^{-8}$ & 1.856 $\pm$ 0.012   & 1.26 (1131/895) \\
	58010.92 $\dagger$ & 10625  & $2.261_{-0.005}^{+0.006}$ & $33.0_{-1.5}^{+1.9}$    & $0.24_{-0.04}^{+0.04}$   & $38.1_{-0.5}^{+0.5}$    &$9.32 \times 10^{-8}$ & 1.798 $\pm$ 0.014   & 1.09 (945/866)  \\
	58011.13 $\dagger$ & 10628  & $2.238_{-0.005}^{+0.005}$ & $21.2_{-0.5}^{+0.6}$    & $0.19_{-0.04}^{+0.04}$   & $31.6_{-1.8}^{+2.2}$    &$9.43 \times 10^{-8}$ & 2.514 $\pm$ 0.012   & 1.19 (1071/895) \\ 
	58011.66 $\dagger$ & 10636  & $2.287_{-0.004}^{+0.009}$ & $41.2_{-2.4}^{+2.9}$    & $0.20_{-0.05}^{+0.02}$   & $46.1_{-0.3}^{+0.2}$    &$1.03 \times 10^{-7}$ & 1.504 $\pm$ 0.010   & 1.34 (1202/895) \\
	58011.87           & 10639  & $2.281_{-0.004}^{+0.005}$ & $36.5_{-2.1}^{+2.5}$    & $0.22_{-0.008}^{+0.006}$ & $31.6_{-6.5}^{+2.4}$    &$7.96 \times 10^{-8}$ & 1.645 $\pm$ 0.009   & 1.29 (351/271)  \\
	58012.24 $\dagger$ & 10644  & $2.327_{-0.005}^{+0.005}$ & $26.1_{-0.8}^{+1.0}$    & $0.22_{-0.003}^{+0.004}$ & $52.0_{-0.4}^{+0.4}$    &$1.03 \times 10^{-7}$ & 2.038 $\pm$ 0.010   & 1.21 (1091/895) \\ 
	58012.67 $\dagger$ & 10651  & $2.316_{-0.005}^{+0.003}$ & $54.4_{-4.4}^{+4.8}$    & $0.19_{-0.01}^{+0.03}$   & $53.9_{-2.5}^{+0.4}$    &$1.10 \times 10^{-7}$ & 1.188 $\pm$ 0.009   & 1.51 (1353/895) \\
	58013.02           & 10656  & $2.249_{-0.004}^{+0.005}$ & $23.6_{-0.58}^{+0.81}$  & $0.29_{-0.005}^{+0.005}$ & $28.4_{-8.6}^{+0.71}$   &$8.48 \times 10^{-8}$ & 2.317 $\pm$ 0.012   & 1.05 (285/271)  \\
	\hline

	\end{tabular}}

\end{table*}

\subsection{Spectral modelling: two-component accretion flow}
\label{sec:tcaf}

Further, we have used the two-component accretion flow
\citep{1995ApJ...455..623C, 2006ApJ...642L..49C} to model the broadband energy
spectra ($0.7 - 80$~keV) using spectral data from {\it SXT} and {\it LAXPC}.
In this model, a Keplerian disc is situated at the equatorial plane of the
binary system and a sub-Keplerian halo component resides on top and bottom of
the Keplerain disc. The inner boundary condition of the accretion flow around
a black hole demands a supersonic flow and the rotating supersonic flow faces a
virtual centrifugal barrier close to the horizon. This triggers a shock
discontinuity in the accretion flow \citep{Fukue1987,Chakra1989} and the flow
becomes sub-sonic after passing through a shock. Here, the kinetic energy gets
converted into thermal energy and this `hot' region beyond the shock is referred
to as the post-shock corona (PSC). The Keplerian disc produces multi-colour
blackbody photons and a significant fraction of these photons intercept the `hot' PSC. The high energy photons are formed due to the inverse Comptonisation of soft photons from the Keplerian disc in the PSC. This model has five parameters \citep[see][]{iyer2015determination} namely the mass of the black hole ($M$), the shock location ($x_s$), the sub-Keplerian halo rate ($\dot m_h$), Keplerian disc accretion rate ($\dot m_d$) and free normalisation constant. Here, mass of the black hole is expressed in solar mass unit ($M_{\odot}$), radial distance is expressed in Schwarzschild radii ($r_g=2GM/c^2$), where $G$ is gravitational constant and $c$ is the speed of light in vacuum. The accretion rate is expressed in unit of Eddington rate ($\dot{m}_{Edd}$).  

\begin{figure}
	\includegraphics[height=\columnwidth,angle=-90]{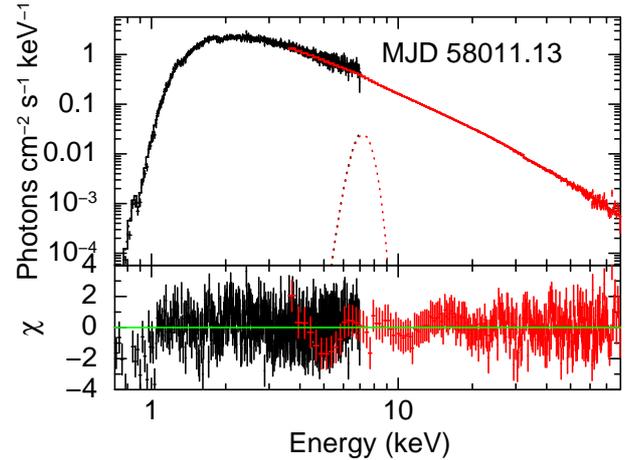}
	\caption{Broadband $0.7 - 80$~keV spectrum for Orbit 10628 (MJD 58011.13) of {\it AstroSat}, using 
	data from both {\it SXT} ($0.7 - 7$~keV) and {\it LAXPC} ($3 - 80$~keV). {\it SXT} data is indicated using black colour and
	{\it LAXPC} data is shown in red colour. The spectrum is fitted with two-component accretion flow   	 
	model. See text for results.}
	\label{fig:broadspec_tcaf}
\end{figure}

This model is implemented in \texttt{XSPEC} as a table model
\citep{iyer2015determination}. In Figure \ref{fig:broadspec_tcaf}, we have
presented a fit of combined spectrum of {\it SXT+LAXPC} data corresponding to
MJD 58011.13 (Orbit number 10628) using two-component model along with an additional
\texttt{highecut} component of \texttt{XSPEC}. The \texttt{gaussian} and \texttt{edge} components used in our earlier model was required here as well. 
The disc and halo accretion rates are
respectively $0.977_{-0.157}^{+0.109}$ and $0.203_{-0.022}^{+0.019}$~$\dot{m}_{Edd}$. The
corresponding shock location is $16.3_{-0.99}^{+0.80}~r_g$. The mass of the black
hole obtained by modelling this particular observation with the two-component
flow model is $6.31_{-0.86}^{+0.68}$ $M_{\odot}$. 
In Table \ref{tab:broadspec_tcaf}, we have presented the results of broadband spectral fitting using two-component accretion flow model (at 90\% confidence level) for two representative cases and observed that the spectral state of the source did not change much. 
Model fitting for the two observations considered suggests that the possible mass of the BH source is in the range $5.14 - 7.83~M_{\odot}$ and the relative accretion rate (see Table \ref{tab:broadspec_tcaf}) indicates the source to be in intermediate state \citep{Nandi2018}. 
We have chosen two broadband ($0.7 - 80$~keV) observations (MJD 58011.13 and 58012.24) for two-component flow modelling as the QPO frequencies in these cases are not the same. We expect that the shock location which is an estimate of the size of the PSC decreases with an increase in QPO frequency. The results of physical modelling and temporal modelling indicates the same, as we have a shock location of $\sim 16.3~r_g$ and QPO frequency of $\sim 2.17$ Hz on MJD 58011.13, while the shock location on MJD 58012.24 is only $\sim 15.9~r_g$ corresponding to a higher QPO frequency of $\sim 2.59$ Hz (see Tables \ref{tab:qpoparAll} and \ref{tab:broadspec_tcaf}). Also, the relative disc to halo accretion rates in both cases are similar, indicating that the source has not undergone a state transition during the {\it AstroSat} observations.

\begin{table}
\caption{Fit parameters from the model \texttt{TBabs} $\times$ \texttt{constant(gaussian} $+$ \texttt{edge} $\times$ \texttt{highecut} $\times$ \texttt{two-component-flow)} applied on {\it AstroSat} data.}
\label{tab:broadspec_tcaf}
\resizebox{\columnwidth}{!}{
\begin{tabular}{|l|l|l|}
\hline
MJD    &  58011.13 &  58012.24           \\
Orbit  & 10628 & 10644                      \\ 
\hline

Mass ($M_{\odot}$)             & $6.31_{-0.86}^{+0.68}$      & $6.47_{-1.33}^{+1.36}$ \\
Shock location ($r_g$)         & $16.3_{-0.99}^{+0.80}$      & $15.9_{-0.8}^{+2.1}$ \\
Halo rate ($\dot{m}_{Edd}$)    & $0.203_{-0.022}^{+0.019}$  & $0.15_{-0.02}^{+0.02}$\\
Disc rate ($\dot{m}_{Edd}$)    & $0.977_{-0.157}^{+0.109}$   & $0.78_{-0.20}^{+0.15}$\\
Cutoff Energy (keV)            & $20.0_{-1.4}^{+1.2}$        & $18.9_{-0.89}^{+1.1}$ \\
Fold Energy (keV)              & $78.6_{-7.8}^{+8.0}$        & $85.2_{-4.7}^{+5.8}$ \\
$\chi ^2/dof$                  & 1.15 (1031.32/891)          & 1.19 (1063.99/891)\\
\hline
\end{tabular}
}
\end{table}

This model has a free normalisation which is a function of inclination angle and distance to the source. The normalisation constant shifts the scale of the overall luminosity of the source. The mass and accretion rates are independent quantities in the model and both determine the source luminosity as well as the spectral characteristics. In fact, for a given mass, the change in accretion rates determines the spectral evolution during the outburst. Hence, only a specific range of mass can model all the observed data. We believe that the observed data may include some contributions from the base of the jet, particularly in intermediate states. This is not considered in the current model. Also, as discussed in \S \ref{sec:disco}, the spin parameter of the black hole is not taken into account in this model. With the inclusion of these contributions in the model, the estimated accretion parameters may differ slightly.

\section{Discussion and Conclusions}
\label{sec:disco}
The {\it MAXI} detection of a bright X-ray transient lead to a large follow up
observation campaign with multiple observatories following the outburst. These
observations point to the black hole binary nature of this source, which shows
state transitions similar to a canonical black hole binary. {\it
AstroSat} made a few observations in the intermediate state of the source. 
In this paper, we have studied the outbursting source MAXI J1535$-$571 using
four different instruments namely {\it MAXI}, {\it Swift/XRT}, {\it AstroSat/SXT} and {\it AstroSat/LAXPC}.
Using {\it MAXI}, we have studied the outburst profile (Figure \ref{fig:fig1}) and the HID (Figure \ref{fig:MAXIXRTHID}) of the
entire outburst. This source showed a slightly unusual {HID} evolution with extremely high
flux seen in the intermediate states before it dimmed considerably during its transition towards the soft state. 
The source has gone through LHS, HIMS, SIMS, HIMS, SIMS and soft state in the rising phase. 
We have not considered data beyond MJD 58200 due to weak activity of the source
leading to large errorbars in counts, though \cite{Tao2018}
have classified the source to be in soft state from  MJD 58137 to 58230. 

We made use of observations from {\it XRT} to 
study the state transitions based on the HRD and RID.
From the HRD (Figure \ref{fig:XRT-HRD-RID}a) it is evident that the hardness decreases as
the states evolve and rms is comparatively lower during
the SIMS. The RID (Figure \ref{fig:XRT-HRD-RID}b) suggests that the 
rms is larger during the
harder states as expected and is low during the SIMS which corresponds to the highest count rate.
The {\it AstroSat} observations of the source shaded in Figures \ref{fig:XRT-HRD-RID}a and \ref{fig:XRT-HRD-RID}b 
falls in the HIMS region.

{\it XRT} and {\it LAXPC} data were used to carry out timing analysis
of the source. We detected C-type QPOs in the range $1.85 - 2.88$~Hz along with second harmonics 
during the rising phase
of the outburst with {\it LAXPC} data. The power spectra also showed band limited noise 
and low frequency noise. We confirm that this period corresponds to the HIMS phase of the
outburst as classified by \cite{Tao2018,2018arXiv180805318H}. 
Using {\it HXMT} data, \cite{2018arXiv180805318H}
have detected QPOs upto 100~keV, though {\it LAXPC} showed no evidence of QPOs beyond
50~keV. This is probably due to the fact that the {\it LAXPC} has
only an effective area in the range 4100 cm$^2$ to 2200 cm$^2$ as energy changes from 60~keV to 80~keV \citep{Antia2017},
whereas {\it HXMT} has a higher effective area of 5100 cm$^2$ in the $20 - 250$~keV energy range \citep{Li2018}. 
QPOs are thought to be signatures of physical oscillations in the accreting flow and 
many reasons have been proposed for explaining the presence of such oscillations. 
Magneto-rotational instabilities (MRIs) in the accretion disc \citep{Machida2008} can be a reason for QPOs.
During the process of accretion, when the magnetic energy accumulated in an accretion disc at a particular stage is released, 
the disc becomes weakly magnetized and axisymmetric. The magnetic energy again gets amplified due to MRI and the cycle repeats.
The period of these cycles results in low frequency peaks around 4 to 8~Hz. LFQPOs in black hole binaries
can possibly be generated due to these magnetic cycles.
Lense-Thirring effects in the accretion disc \citep{Ingram2009} may also be a reason for QPOs. In this case,
the rotation of the central black hole causes frame-dragging in the vicinity of the black hole. This in turn, causes the orbits of
particles in the accretion disc to precess and the associated precession frequency is found to be close to
observed QPO frequencies.
In this work, we consider the two-component accretion flow framework, 
under which an oscillating PSC \citep{Chakra2008} can cause these QPOs.

It is evident from the energy dependent analysis of power spectra that the QPO rms exceeded $\sim$8\% in $3 - 20$~keV energy band. 
It increased up to $\sim$9\% in the $20 - 35$~keV range and in the $35 - 50$~keV energy band the QPO rms decreased slightly to $6 - 7$\%. \cite{Rodriguez2002} have reported similar energy dependence of LFQPOs for GRS 1915+105. \cite{Rao2016} have studied the connection between the QPOs and the spectral components and have shown that a small in-phase variation of the spectral index with the QPO could reproduce the energy dependence of QPO rms. This is consistent with the origin of LFQPOs due to oscillation of PSC. For a given QPO frequency, the PSC will oscillate with respect to an average position. During oscillation, as PSC expands the spectral index decreases (i.e. harder spectrum) and when the PSC shrinks the spectral index increases. This behaviour is same as the change in spectral index due to the change in the shock location as shown in \cite{MC2005}.

Further, {\it SXT+LAXPC} combined broadband spectra
were analysed to examine the source characteristics. Initially, the spectra were modelled
with a thermally Comptonised continuum (\texttt{nthComp}) along with a Xenon \texttt{edge} and an iron line (\texttt{gaussian}).
The seed photon temperature was consistently in $0.19$ to $0.29$~keV range. The high energy cut off
was at least $\sim 60$~keV as inferred from the electron temperature ($kT_e$).  
The photon index
of the energy spectrum was in a narrow range of $2.18 - 2.37$ throughout
the observations with {\it AstroSat}. This along with the lack of QPO frequency
evolution indicates that the oscillating post shock corona was almost constant in size during this period. 
The phenomenological modelling implies that the
spectrum was completely dominated by thermal Comptonisation. The absence of a separate
disc component in the model, suggests that disc was either truncated away as indicated by the low seed
photon temperature or was buried within the corona. 
The soft photons supplied by the disc
enters into the corona which has a moderate optical depth ($0.72 < \tau < 2.5$) and produces
the strong comptonised component.

Moreover, as the relative reflection coefficient was not significant in the energy spectra of HIMS observations with {\it AstroSat}, we expected that the iron abundance must be high to get a strong iron line feature. In order to verify the results, we modelled the {\it SXT}~+~{\it LAXPC} broadband spectra of HIMS with \texttt{relxillCp} \citep{Garcia2014}, which includes Comptonisation and reflection taking into consideration the iron abundance. The value of reflection coefficient obtained from \texttt{relxillCp} was $0.07 \pm 0.02$ and the iron abundance was in the range between $1.2 - 1.7$ times solar abundance. The larger than solar abundance levels of iron caused the strong iron lines in the energy spectra. The iron abundance is comparable to the value $1.4_{-0.1}^{+0.3}$ times solar abundance, as obtained by \cite{2018ApJ...852L..34X}, during LHS (MJD 58003). The equivalent width (EW) of the iron line was $\sim 0.11$ keV, which falls in the range between 0.05~keV to 0.3~keV, typical for black hole binaries \citep{Gilfanov2010}.

Besides this, we modelled the combined {\it SXT + LAXPC} energy spectra using the two-component accretion flow model.
The results from
Table~\ref{tab:broadspec_tcaf} show that the Keplerian-disc accretion rate is
around the Eddington accretion limit.
Similar rates are seen in XTE J1859$+$226 
\citep{Nandi2018} during its HIMS, but not usually seen in other black hole binaries
\citep[see][]{iyer2015determination, Rad2018}. 
The near Eddington accretion rates (see Table \ref{tab:broadspec_tcaf}) observed in the
intermediate states seem to be the reason for the excessive brightening of the source with luminosity as high as 1.07 $L_{Edd}$ (see Table \ref{tab:specpar} and \S \ref{sec:spec}). 
High accretion rates can trigger outflows to escape from the accretion
stream, which can be observationally seen as radio emissions \citep{radnan,Rad2016}.
The large absorption columns
reported (as compared to the expected galactic column from HI maps)
also point to the presence of some outflows. \cite{baglio18} report
wild infra-red flickering during this period, consistent with erratic outflows.

The spectral modelling with two-component flow suggests a compact corona of almost constant size ($\sim 16~r_g$).
As a result, the C-type QPOs formed due to the oscillations of the corona are in a narrow range of frequencies as observed from {\it LAXPC} data. 
It should be noted that, the model is applicable only for a stationary black hole. The effect of black hole spin on the radiation spectrum is that the spectrum gets harder \citep{Mandal2018} as the black hole spins faster. Hence, inclusion of black hole spin will change the estimated value of the size of PSC for a given data set. In future, we aim to model the spectral data including the black hole spin.

Finally, physical modelling of the energy spectra from {\it AstroSat} observations based on the two-component accretion flow model suggests
that the source hosts a black hole of mass $5.14 - 7.83$~$M_{\odot}$, which is similar to
other Galactic black hole binaries \citep{Corral2016,Tetarenko2016,Nandi2018}. 
In summary, we have characterised the spectro-temporal properties of the source MAXI J1535$-$571 using {\it AstroSat} data during its 2017 outburst with the help of {\it MAXI} and {\it Swift} results.

\section*{Acknowledgement}
Authors are thankful to the reviewer for his/her
valuable suggestions and comments that helped to improve the quality of the manuscript.
We acknowledge the clarifications that we got from Dr. Matteo Bachetti regarding dead-time effects. 
We are thankful to the {\it Swift} help desk for the information on pile-up and dead-time effects in {\it XRT}.
This research has made use of data obtained
through the High Energy Astrophysics Science Archive Research Center (HEASARC)
online service, provided by the NASA/Goddard Space
Flight Center. 
This publication uses the data from the {\it AstroSat} mission of the Indian Space
Research Organisation (ISRO), archived at the Indian Space Science Data
Centre (ISSDC).
This work has used the data from the Soft X-ray Telescope ({\it SXT}) developed at TIFR,
Mumbai, and the {\it SXT} POC at TIFR is thanked for verifying \& releasing the data and providing the necessary software tools. 
This work has also used the data from the {\it LAXPC} Instruments developed at TIFR, Mumbai, and the {\it LAXPC} POC 
at TIFR is thanked for verifying \& releasing the data. 
We thank the {\it AstroSat} Science Support Cell hosted by IUCAA and TIFR for providing the \texttt{LaxpcSoft} software 
which we used for {\it LAXPC} data analysis. 
AN thanks GD, SAG; DD, PDMSA and Director, URSC
for encouragement and continuous support to carry out this research.




\bibliographystyle{mnras}
\bibliography{maxij1535-571}{}\label{bib}



\clearpage
\appendix

%

\section{Observations with {\it AstroSat}}

\begin{table}

  \begin{minipage}{0.5\textwidth}
	\centering
	\caption{Observations with {\it AstroSat}: ObsID T01\_191T01\_9000001536
	between $12^{\mathrm{th}}$~Sep till $17^{\mathrm{th}}$~Sep, 2017.
	Fifteen orbits chosen are bold-faced.} 
	\label{tab:astrosatobsall}
	\scriptsize{\begin{tabular}{|c|c|l|c|c|}
	\hline
		Orb-\#  &	Exp Time (s) &	Start Time (UTC) & End Time (UTC)\\
	\hline \hline

    10584 &	    496.953770123 &	2017-09-12 05:32:33 	&2017-09-12 06:04:44 \\ 
    \bf{10585} &	    4941.77365309 &	2017-09-12 05:55:48 	&2017-09-12 07:50:36 \\ 
    10586 &	    5357.744978 	&2017-09-12 07:35:51 	&2017-09-12 09:36:54 \\ 
    10588 &	    7382.02297796 &	2017-09-12 09:25:32 	&2017-09-12 12:50:29 \\ 
    \bf{10589} &	    4583.00317103 &	2017-09-12 12:45:50 	&2017-09-12 14:49:28 \\ 
    10590 &	    4031.87576006 &	2017-09-12 14:36:24 &	2017-09-12 16:33:09 \\ 
    10592 &	    3851.26237635 &	2017-09-12 16:21:05 &	2017-09-12 18:16:50 \\ 
    10593 &	    3531.06603971 &	2017-09-12 18:08:39 	&2017-09-12 20:00:38 \\ 
    \bf{10594} &	    3292.33470635 &	2017-09-12 19:50:12 	&2017-09-12 21:43:51 \\ 
    10595 &	    2970.22884139 &	2017-09-12 21:33:21 &	2017-09-12 23:28:27 \\ 
    10596 &	    3249.72885065 &	2017-09-12 22:59:54 	&2017-09-13 01:12:44 \\ 
    \bf{10597} &	    3264.45310393 &	2017-09-13 00:39:54 	&2017-09-13 03:02:45 \\ 
    10598 &	    4230.74553183 &	2017-09-13 02:39:27 	&2017-09-13 04:45:33 \\ 
    10599 &	    4552.78409085 &	2017-09-13 04:28:08 	&2017-09-13 06:28:26 \\ 
    10600 &	    5069.68189324 &	2017-09-13 06:13:33 	&2017-09-13 08:13:57 \\ 
    10603 &	    12317.7855291 &	2017-09-13 08:06:35 	&2017-09-13 13:29:21 \\ 
    \bf{10604} &	    3694.92741311 &	2017-09-13 13:21:28 	&2017-09-13 15:11:56 \\ 
    10606 &	    3628.15777545 &	2017-09-13 15:04:25 	&2017-09-13 16:55:33 \\ 
    10607 &	    3874.79451605 &	2017-09-13 16:42:45 	&2017-09-13 18:39:45 \\ 
    \bf{10608} &	    3442.76126461 &	2017-09-13 18:31:27 	&2017-09-13 20:23:45 \\ 
    10609 &	    3070.35393815 &	2017-09-13 20:14:13 	&2017-09-13 22:05:06 \\ 
    10610 &	    3307.57441911 &	2017-09-13 21:48:02 	&2017-09-13 23:50:57 \\ 
    10611 &	    3153.40418664 &	2017-09-13 23:31:51 	&2017-09-14 01:35:17 \\ 
    \bf{10612} &	    4134.9651357 &	2017-09-14 00:46:03 	&2017-09-14 03:25:09 \\ 
    10613 &	    4389.06476232 &	2017-09-14 03:03:45 	&2017-09-14 05:08:50 \\ 
    10614 &	    4761.01065524 &	2017-09-14 04:47:38 	&2017-09-14 06:52:10 \\ 
    10615 &	    5219.04197027 &	2017-09-14 06:39:20 	&2017-09-14 08:38:08 \\ 
    10617 &	    8634.5027135 	&2017-09-14 08:25:20 &	2017-09-14 12:07:52 \\ 
    \bf{10618} &	    4120.10051683 &	2017-09-14 11:56:09 &	2017-09-14 13:51:28 \\ 
    10619 &	    3991.58977488 &	2017-09-14 13:39:03 	&2017-09-14 15:34:31 \\ 
    10621 &	    3821.43898252 &	2017-09-14 15:23:18 	&2017-09-14 17:18:31 \\ 
    10622 &	    3667.28994137 &	2017-09-14 17:09:19 	&2017-09-14 19:02:57 \\ 
    10623 &	    3609.30743047 &	2017-09-14 18:49:47 	&2017-09-14 20:47:03 \\ 
    10624 &	    3379.89339639 &	2017-09-14 20:31:08 	&2017-09-14 22:30:13 \\ 
    \bf{10625} &	    3406.84135797 &	2017-09-14 22:09:55 	&2017-09-15 00:14:39 \\ 
    10626 &	    3357.2935165 	&2017-09-14 23:43:14 &	2017-09-15 01:59:46 \\ 
    10627 &	    4507.45603695 &	2017-09-15 01:13:12 &	2017-09-15 03:48:27 \\ 
    \bf{10628} &	    4542.56313907 &	2017-09-15 03:10:26 &	2017-09-15 05:31:16 \\ 
    10629 &	    4904.77152288 &	2017-09-15 05:22:25 	&2017-09-15 07:15:57 \\ 
    10632 &	    12804.8367192 &	2017-09-15 07:02:51 &	2017-09-15 12:31:09 \\ 
    10633 &	    4078.40076538 &	2017-09-15 12:18:56 	&2017-09-15 14:14:48 \\ 
    10635 &	    3657.04443797 &	2017-09-15 14:06:59 	&2017-09-15 15:57:55 \\ 
    \bf{10636} &	    3488.04376683 &	2017-09-15 15:50:31 	&2017-09-15 17:41:02 \\ 
    10637 &	    3765.87370964 &	2017-09-15 17:28:37 &	2017-09-15 19:25:07 \\ 
    10638 &	    3336.03761904 &	2017-09-15 19:15:16 &	2017-09-15 21:08:22 \\ 
    \bf{10639} &	    3148.43251339 &	2017-09-15 20:56:25 &	2017-09-15 22:53:34 \\ 
    10640 &	    3081.41305334 &	2017-09-15 22:37:55 	&2017-09-16 00:38:52 \\ 
    10641 &	    3561.31836302 &	2017-09-16 00:03:14 	&2017-09-16 02:22:36 \\ 
    10642 &	    4247.31112233 &	2017-09-16 01:51:34 	&2017-09-16 04:11:31 \\ 
    10643 &	    4544.92289703 &	2017-09-16 03:54:46 	&2017-09-16 05:54:00 \\ 
    \bf{10644} &	    4903.1540284 &	2017-09-16 05:46:24 	&2017-09-16 07:40:10 \\ 
    10646 &	    8637.82817281 &	2017-09-16 07:33:24 	&2017-09-16 11:09:57 \\ 
    10647 &	    3889.49490765 &	2017-09-16 11:03:05 &	2017-09-16 12:54:18 \\ 
    10648 &	    3992.12343246 &	2017-09-16 12:41:34 &	2017-09-16 14:37:30 \\ 
    10650 &	    3518.57945908 &	2017-09-16 14:29:53 	&2017-09-16 16:20:18 \\ 
    \bf{10651} &	    3424.59817114 &	2017-09-16 16:13:03 &	2017-09-16 18:04:24 \\ 
    10652 &	    3822.22094573 &	2017-09-16 17:50:32 	&2017-09-16 19:48:56 \\ 
    10653 &	    3282.38882317 &	2017-09-16 19:36:53 	&2017-09-16 21:30:31 \\ 
    10654 &	    3198.43616311 &	2017-09-16 21:16:46 &	2017-09-16 23:15:04 \\ 
    10655 &	    3446.71049657 &	2017-09-16 22:41:34& 	2017-09-17 01:00:46 \\ 
   \bf{ 10656} &	     3721.80164763 &	2017-09-17 00:29:54 	&2017-09-17 02:46:05 \\ 
	10657 &			2967.15269032 &	2017-09-17 02:21:23 	&2017-09-17 03:42:35 \\ 

	\hline

	\end{tabular}}

  \end{minipage}

\end{table}

\section{Observations with {\it Swift}}

\begin{table}

  \begin{minipage}{0.4\textwidth}
	\centering
	\caption{{\it Swift} Public archive. Selected datasets used in
         our analysis have bold-faced ObsIds.} 
	\label{tab:swiftobsall}
	\scriptsize{\begin{tabular}{|c|c|c|}
	\hline
		ObsId &	Start Time (UTC) & XRT Exp (s) \\
	\hline \hline
                00770431000	&	2017-09-02 19:44:52	&	1615.85\\
            \bf{00770431001}	&	2017-09-02 22:11:56	&	3387.90		\\
                00770502000	&	2017-09-03 08:44:02	&	75.03		\\
                00770656000	&	2017-09-04 13:29:33	&	15.59		\\
            \bf{00770656001}	&	2017-09-04 17:10:06	&	988.79			\\
            \bf{00770656002}	&	2017-09-05 05:17:20	&	4504.50			\\
                00010264002	&	2017-09-06 00:38:57	&	449.22		\\
            \bf{00010264003}	&	2017-09-08 06:38:57	&	4948.52			\\
                00771371000	&	2017-09-08 07:58:07	&	69.12			\\
            \bf{00010264004}	&	2017-09-11 06:32:57	&	1089.60		\\
            \bf{00010264005}	&	2017-09-12 06:16:57	&	899.60	    \\
            \bf{00010264007}	&	2017-09-13 00:08:57	&	889.45		\\
            \bf{00010264006}	&	2017-09-14 22:17:57	&	1074.60		\\
            \bf{00010264008}	&	2017-09-15 09:15:57	&	984.63		\\
            \bf{00010264009}	&	2017-09-16 02:44:57	&	1059.78			\\
            \bf{00088245001}	&	2017-09-17 04:14:57	&	2223.08		\\
            \bf{00010264010}	&	2017-09-18 04:10:56	&	1104.60		\\
            \bf{00010264011}	&	2017-09-19 08:52:57	&	989.38			\\
                00088245002	&	2017-09-20 23:10:57	&	954.61		\\
            \bf{00088245003}	&	2017-09-21 00:46:57	&	1014.18		\\
                00010264012	&	2017-09-22 15:15:57	&	1189.20			\\
                00010264013	&	2017-09-23 02:20:56	&	1009.60			\\
                00010264014	&	2017-09-24 02:25:57	&	599.20		\\
                00773893000	&	2017-09-24 09:54:14	&	620.93			\\
            \bf{00010264015}	&	2017-09-25 21:28:56	&	899.48		\\
            \bf{00010264016}	&	2017-09-26 18:12:56	&	909.21		\\
            \bf{00010264017}	&	2017-09-27 16:37:57	&	859.22		\\
            \bf{00010264018}	&	2017-09-28 18:01:57	&	1044.60			\\
            \bf{00010264019}	&	2017-09-29 17:56:57	&	964.65		\\
            \bf{00010264020}	&	2017-09-30 17:56:57	&	889.61			\\
            \bf{00010264021}	&	2017-10-01 17:46:56	&	874.61		\\
            \bf{00010264022}	&	2017-10-02 17:44:56	&	1109.60		\\
            \bf{00010264023}	&	2017-10-03 17:44:22	&	1004.60		\\
            \bf{00010264024}	&	2017-10-04 17:40:57	&	964.60		\\
                00010264025	&	2017-10-05 15:53:27	&	604.61	\\
            \bf{00010264026}	&	2017-10-06 11:06:57	&	1154.60			\\
            \bf{00010264027}	&	2017-10-07 04:35:57	&	1024.60			\\
            \bf{00010264028}	&	2017-10-08 05:53:56	&	974.60		\\
            \bf{00010264029}	&	2017-10-09 10:51:57	&	1059.61			\\
            \bf{00010264030}	&	2017-10-10 12:03:57	&	994.06			\\
            \bf{00010264031}	&	2017-10-11 07:32:57	&	854.62			\\
            \bf{00088245004}	&	2017-10-22 23:40:57	&	2088.03		\\
            \bf{00088246001}	&	2017-10-24 23:28:57	&	1528.19			\\

		00010490007	&	2017-12-31 06:36:57	&	109.75		\\
		00010491001	&	2017-12-31 06:39:31	&	869.95		\\
		00010492001	&	2017-12-31 06:41:06	&	864.96		\\
    \bf{00010493001}	&	2017-12-31 06:42:39	&	915.76		\\
		00010494001	&	2017-12-31 06:44:12	&	877.77		\\
		00010495001	&	2017-12-31 06:45:41	&	885.02	\\
		00010496001	&	2017-12-31 06:47:05	&	724.70	\\
		00010497001	&	2017-12-31 06:48:28	&	847.41		\\
	\bf{00010491002}	&	2018-01-19 08:05:57	&	1119.61	\\
		00010264032	&	2018-03-17 19:13:28	&	279.61		\\
	\bf{00010264033}	&	2018-03-18 10:47:57	&	1370.34	\\
		00010264034	&	2018-03-24 21:16:57	&	1859.22		\\
		00010264035	&	2018-03-26 13:28:47	&	1409.22	\\
		00010264036	&	2018-03-28 13:23:57	&	2015.28		\\
		00010264037	&	2018-03-30 02:03:57	&	1910.27		\\

	\hline

	\end{tabular}}

  \end{minipage}

\end{table}

\bsp	
\label{lastpage}
\end{document}